\documentclass[11pt,english,a4paper,floatfix]{article}
\usepackage{jcappub}
\usepackage{multirow}
\usepackage[normalem]{ulem}
\pdfoutput=1
\usepackage{bm}
\usepackage{amsfonts}
\usepackage[utf8]{inputenc}
\graphicspath{{Figures/}}
\newcommand{\Hunit}{km s$^{-1}$ Mpc$^{-1}$}

\usepackage[dvipsnames]{xcolor}
\usepackage{graphicx}
\usepackage{amssymb}
\usepackage{amsmath}
\usepackage{times}
\usepackage{multirow}
\usepackage{url}
\usepackage{cuted}
\usepackage[english]{babel}
\usepackage{enumerate}
\usepackage[normalem]{ulem}
\usepackage{enumitem}
\usepackage{multirow}
\usepackage{float}
\usepackage{booktabs}
\usepackage{makecell}
\usepackage{placeins}
\usepackage{newtxtext,newtxmath}
\usepackage[T1]{fontenc}
\usepackage{amsmath}	
\usepackage{csquotes}
\usepackage[dvipsnames]{xcolor}
\usepackage{bigints}
\newcommand{\Rsix}{SH0ES'16 }

\newcommand{\Rtwo}{SH0ES'22 }

\begin{document}
	
	
	\title{A gravitational constant transition within cepheids as supernovae calibrators can solve the Hubble tension}
	
	\author[a,b]{Ruchika,}
	\author[a,c]{Himansh Rathore,}
	\author[d,e]{Shouvik Roy Choudhury,}
	\author[a]{Vikram Rentala}
	
	\affiliation[a]{Department of Physics, Indian Institute of Technology Bombay,
		Main Gate Road, Powai, Mumbai, Maharashtra 400076, India}
	\affiliation[b]{Istituto Nazionale di Fisica Nucleare (INFN), Sezione di Roma, P.le A. Moro 2, I-00185, Roma, Italy}	
	\affiliation[c]{Department of Astronomy \& Steward Observatory, University of Arizona, 933 N. Cherry Avenue, Tucson, Arizona 85721, USA}	
	\affiliation[d]{Inter-University Centre for Astronomy \& Astrophysics (IUCAA), Ganeshkhind, Post Bag 4, Pune, Maharashtra 411007, India}
	\affiliation[e]{Institute of Astronomy and Astrophysics, Academia Sinica, No. 1, Sec. 4, Roosevelt Road, Taipei 10617, Taiwan}
	\emailAdd{ruchika.ruchika@roma1.infn.it}
	\emailAdd{himansh@arizona.edu}
	\emailAdd{sroy@asiaa.sinica.edu.tw}
    \emailAdd{rentala@phy.iitb.ac.in}


\abstract{
Local universe measurements of the Hubble constant $H_0$ using SNe~Ia with Cepheids as calibrators yield a value of $H_0$ = 73.04 $\pm$ 1.04 \Hunit, which is in tension with the value of $H_0$ inferred from the Cosmic Microwave Background and other higher redshift probes.
In ref.~\cite{Marra:2021fvf}, the authors proposed a rapid transition in the value of the effective Newtonian gravitational constant $G$ in order to alleviate the Hubble tension. The transition point was chosen so as to only affect distance estimates to Hubble flow SNe. However, in this study, the authors made the assumption that SNe~Ia peak luminosity $L$ increases with Chandrasekhar mass $M_c$. This hypothesis contradicts a previous semi-analytic study of SN light curves in the presence of a $G$-transition~\cite{Wright2018} which concluded that there is an inverse relationship between $L$ and $M_c$. Motivated by the results of ref.~\cite{Marra:2021fvf} and~\cite{Wright2018}, we propose a hypothesis of a sudden recent change in the effective Newtonian gravitational constant $G$ at an epoch which corresponds to a smaller look-back distance between $\sim$~7~-~80~Mpc. A transition in $G$ at these distances would affect both our estimates of the distances to Cepheids in calibrator galaxies, as well as to the Hubble flow supernovae. Upon fitting the observational data to this hypothesis, we find three interesting results: (i) we find mild evidence for a $G$-transition at 22.4 Mpc (73 million years ago)  which is preferred (using certain estimators) by the calibrator type Ia SNe data over no $G$-transition, (ii) the Hubble constant inferred under this hypothesis is in good agreement with the value obtained from the CMB for a 4\% larger value of $G$ at earlier times, thus potentially resolving the Hubble tension, (iii) we obtain a fit to the scaling relationship between SN peak luminosity $L$ and Chandrasekhar mass $M_c$, as $L\propto M_c^{-1.68 \pm 0.68}$, which is in good agreement with the prediction of the theoretical study of ref.~\citep{Wright2018}. We also discuss how other probes could be used to verify this transition in the value of $G$.}

\maketitle



\section{Introduction}
The measurement of the Hubble constant $H_0$ is of fundamental importance to calibrate our cosmological models. $H_0$ can be inferred either directly from low-redshift probes of the expansion rate of the local universe, or it can be inferred indirectly through measurements of high redshift observables such as the Cosmic Microwave Background (CMB) and Baryon Acoustic Oscillations (BAO). The local measurement of $H_0$ from the Cepheid-calibrated type Ia Supernovae (SNe Ia), as performed by the \Rtwo collaboration \cite{R22} yields a value of $H_0 = 73.04 \, \pm \, 1.04$ \Hunit (68\% C.L.). In contrast, the high redshift measurements from CMB and BAO yield a lower value of $H_0 = 67.66 \, \pm  \, 0.42$~\Hunit (68\% C.L.) \cite{p2020}. These two values are discrepant at the  5-$\sigma$ level. This discrepancy between these estimated values of the Hubble constant $H_0$ from the local universe and CMB measurements is called the ``Hubble tension''. Resolving this discrepancy is one of the major open problems in cosmology.

The low-redshift probes rely on the assumption of a ``standard candle'' which are type Ia supernovae. The distances to these objects can not be calibrated by a direct method such as parallax measurements, and therefore another intermediate calibrator is needed. Cepheid variables, which have been shown to have a robust period-luminosity relationship (PLR) and can be seen out to large distances, have been used as calibrators in the SH0ES analysis~\cite{R22}.

It is possible that unknown systematics in the physics of Cepheid calibrators could perhaps be responsible for the discrepancy in the inferred value of $H_0$. The Carnegie-Chicago Hubble Project (CCHP) collaboration has attempted to measure the $H_0$ parameter using SNe Ia calibrated to stars at the tip of the red-giant branch (TRGB), and they obtained a value of $H_0 = 69.8\pm0.8~(\textrm{stat})\pm1.7~(\textrm{sys})$~\Hunit (68\% C.L.) \cite{Freedman:2019jwv, Freedman:2021ahq}, which lies between the values reported by SH0ES \citep{R19,R21,R22, R16} and the Planck 2018 results \citep{p2020} and can be taken to agree with either within 2$\sigma$. However, a revised calibration of TRGBs using the parallax measurement of $\omega-$Centauri from GAIA EDR3 yields a value of $H_0 = 72.1\pm2.0$ \Hunit (68\% C.L.) \cite{Yuan:2019npk,Soltis:2020gpl}, which makes both the local CCHP measurement and SH0ES measurement consistent with each other, while still indicating a tension between the low-redshift and high-redshift measurements.

Other local measurements from quasar strong lensing, dispersion of fast radio burst signals, and gravitational wave observations have not yet achieved the precision needed to weigh in on the discrepancy. A joint analysis of $6$ strongly lensed quasars with measured time delays yielded~$H_0 = 73.3^{+1.7}_{-1.8}$~\Hunit \citep{Wong}. An analysis of a set of $9$ currently available Fast Radio Burst (FRB) \citep{9frb} samples yielded $H_0 = 62.3\pm9.1$~\Hunit. Recent gravitational wave events with the first and second observing runs at the advanced LIGO/Virgo along with binary black hole detections in conjunction with galaxy catalogs have found $H_0 = 68.7^{+17.0}_{-7.8}$~\Hunit \citep{LIGOScientific}.

A number of studies have been conducted to understand if as yet unknown systematic effects in the measurements of the local universe are the cause of the Hubble tension \cite{Mortsell:2021nzg,Mortsell:2021tcx,Efstathiou:2020wxn,Cerny:2020inj,Rigault:2014kaa,NearbySupernovaFactory:2018qkd,Jones:2018vbn,Brout:2020msh,Kenworthy:2019qwq}. However, accounting for these systematic effects does not seem to resolve the tension.

Given that the Hubble tension is unlikely to be resolved with standard physics alone, a variety of new physics solutions have been proposed that attempt to resolve this tension (see \cite{DiValentino:2021izs} and the references therein). These solutions can, broadly, be classified into two types, depending on where the new physics has its strongest effect, as pre-recombination and post-recombination solutions~\cite{Knox:2019rjx}. While some of the proposed solutions ameliorate the Hubble tension, there are no known solutions that can fully resolve it (i.e. reduce the discrepancy to less than 1$\sigma$ without creating other discrepancies)~\cite{Schoneberg:2021qvd}.

In this work, we focus on a sharp transition in the gravitational constant ($G$) in the very late universe, as a potential solution to the $H_0$ tension. Such a late transition in the gravitational constant would, among other things, change the physics of Cepheids and SNe~Ia, thus modifying our inferred distance measurements, and hence the value of the inferred Hubble constant.

Our motivation to study the effects of such a transition stems from a series of other works which have discussed the possibility of a $G$-transition in the late universe.
\begin{itemize}
\item In \cite{Desmond:2019ygn} and \cite{Desmond:2020wep}, the authors discussed how changes in the effective gravitational constant due to screened fifth forces can cause changes in the dynamics of Cepheids and TRGB stars respectively, which can potentially solve the Hubble tension. However, in this work, the authors did not assume a sharp transition in $G$, but rather a local environmental dependence on $G$ using the fifth force mechanism. They could alleviate the Hubble tension to below $3\sigma$ but not less than 2$\sigma$ while maintaining self-consistency of the distance ladder.

\item In \cite{Marra:2021fvf}, the authors argued how the Hubble tension can be interpreted as a tension in the inferred absolute magnitude ($M_B$) of distant SNe~Ia which lie in the Hubble flow. Such an effect could arise due to a rapid transition in the gravitational constant at a certain transition redshift\footnote{In this study the authors took the transition to occur at $z\simeq 0.01$, which corresponds to distances $\gtrsim 40$ Mpc in the standard cosmology. Subsequent to this work, calibrator galaxies have been used out to 80 Mpc in \Rtwo, and thus the work of these authors can be reinterpreted as a transition at 80 Mpc, such that once again only Hubble flow SNe are affected.}.

\item The possibility of a sharp transition in the gravitational constant was also discussed as an explanation of sudden transitions in the evolution of the Tully-Fisher relation in \cite{Alestas:2021nmi}. They found a shift or transition in the intercept of the logarithmic form of BTFR (Baryonic Tully-Fisher relation) at a transition distance ($d_T$) of either $9$ or $17$ Mpc.
\end{itemize}

In the present work, we study the possibility of a late-time $G$-transition at look-back times corresponding to a distance between $\sim$ 7 - 80 Mpc (23 million to 260 million years ago). Such a late transition would not only affect the physics of SNe~Ia, but it would also alter the inferred distances to standard calibrators such as Cepheids and TRGB stars. Thus, our hypothesis of new physics is distinct from the work of~\cite{Marra:2021fvf} which considered a change which can only affect type Ia SNe. Our hypothesis is also distinct from that of \cite{Desmond:2019ygn} and \cite{Desmond:2020wep} which only considered a change in the physics of Cepheids and TRGBs due to an environmental $G$ dependence. In addition to these distinctions, a unique feature of our work is that we also take care to include the implications of a $G$-transition on cosmological scales and the consequent implications for extraction of cosmological observables.

While the effect of a $G$-transition on the Cepheid PLR can be easily modelled and used to recalibrate the distance ladder to Cepheids, the effect of a $G$-transition on the luminosity of type~Ia supernovae is more uncertain. The key effect of a $G$-transition here is that it would change the Chandrashekhar mass $M_c$, and type~Ia SNe luminosities are assumed to grow with $M_c$. However, the precise form of the scaling relation is unknown. Wright and Li~\cite{Wright2018} in a theoretical study with non-standard gravity argued that the standardized SN peak luminosity \textit{decreases} with an increase in the Chandrasekhar Mass $M_c$ rather than increasing. They found that the scaling relation of the standardized SN luminosity with $M_c$ is $L\sim M_c^{-0.97}$~\cite{Sakstein2019}.

We perform a fit similar in spirit to that of the SH0ES collaboration \cite{R21} to Cepheids and SNe~Ia observational data, but under the modified hypothesis of a late-time $G$-transition. Rather than assuming a specific scaling relation of $L$ with $M_c$, we parameterize the scaling relation of type Ia SNe as $L\sim M_c^{n}$, and we leave $n$ as a fit parameter. Upon fitting the low redshift data to this hypothesis, we find three interesting results -- (i) we find mild evidence that the $G$-transition that we propose is preferred by the type Ia SNe data over no $G$-transition, (ii) the Hubble constant inferred under this hypothesis is in good agreement with the value obtained from the CMB for a 4\% larger value of $G$ at earlier times, thus potentially resolving the Hubble tension, (iii) we obtain a fit to the scaling relationship between SN peak luminosity $L$ and Chandrasekhar mass $M_c$, as $L\propto M_c^{-1.68 \pm 0.68}$, which is in agreement with the prediction of the theoretical study of ref.~\citep{Wright2018}.

Our results suggest circumstantial evidence for a late time $G$-transition as a solution to the Hubble tension.  We also discuss further tests that could be performed to confirm, or rule out the $G$-transition hypothesis.

How can such a $G$-transition arise in a fundamental theory? In principle, scalar-tensor theories of gravity are well known examples of scenarios where the effective gravitational coupling can acquire a spacetime dependence. Such theories are however subject to stringent constraints from solar-system tests of general relativity~\cite{Uzan:2010pm}. Arranging for a sharp $G$-transition in such models, while at the same time being consistent with these constraints, is a challenge for model building of such a scenario. In this work we take a phenomenological approach and simply assume that such a model can be constructed. However, we will lay out our working assumptions of such a model so that it is clear what effect we expect from such a transition on our cosmological parameter inferences.

This paper is structured as follows. In section~\ref{sec:assumptions}, we lay out the assumptions of our cosmological scenario involving a late-time $G$-transition. In section~\ref{sec:CMBinference}, we explain how \textit{there are not one but two candidate parameters for the Hubble constant} in such a scenario, which we dub $H_{0\alpha}$ and $H_{0\beta}$. We argue that the CMB based extraction for the parameter $H_0$ in the standard cosmology can be reinterpreted as an extraction of the constant $H_{0\beta}$ of the modified theory. We also argue that it is the effective parameter $H_{0\beta}$ that is being measured by direct distance ladder probes at low redshifts, but here a reanalysis of the data is required in order to correctly extract the value of $H_{0\beta}$. In section~\ref{sec:intuition}, we give analytic arguments for the effect of a $G$-transition on the inference of the Hubble constant using low redshift probes, through the effects on the Cepheid period-luminosity relation~(PLR), the SNe Ia luminosity, and the Hubble flow SNe distance-redshift relation. We then explain the consequent effect on distance ladder inferences and inference of the Hubble constant if one were to incorrectly assume that no $G$-transition had occurred, and we argue that this can explain why the low redshift probes give a larger value of the Hubble constant.

In the rest of the paper, we will proceed to reanalyze the low redshift data to extract the constant $H_{0\beta}$ after correctly accounting for a $G$-transition. In section~\ref{sec:analysis}, we outline the methodology that we will use in performing our fit to the distance ladder in the presence of a $G$-transition. Then in section~\ref{sec:data}, we discuss the observational data sets of Cepheids and supernovae used for fitting the distance ladder. In section \ref{sec:standardfit}, we apply a simplified analysis procedure to reproduce the results of the \Rtwo~\cite{R22} fit to the distance ladder, which assumes the standard cosmology without a $G$-transition. After validating this analysis strategy, we then proceed in section~\ref{sec:Gtransitionfit} to discuss the change to our analysis method that is needed when taking into account the possibility of a $G$-transition. In the same section, we also show the results of our analysis when including a $G$-transition. We demonstrate two of our main claims that we have stated above, about resolving the Hubble tension and our inference of the $L$-$M_c$ relation in this section. In section \ref{sec:comparison}, we use different fit comparison techniques like $\chi^2_{\textrm{dof}}$, AIC, and BIC to understand the preference in the data for a $G$-transition hypothesis over the null hypothesis of the no $G$-transition scenario. We finally conclude with some discussion on implications of our results and further tests in section~\ref{sec:discussion}.

In appendix~\ref{sec:scalartensor}, we discuss the challenges of constructing a scalar-tensor theory that can give rise to a $G$-transition and how such a theory motivates our working assumptions. In appendix~\ref{sec:appendixB}, we calculate the change in the value of the Hubble constant inferred from the CMB assuming a $G$-transition.

\section{Laying out the assumptions of our $G$-transition hypothesis}
\label{sec:assumptions}
In this section we lay out our assumptions of what it means for the gravitational constant $G$ to undergo a transition. Since a change in $G$ affects physics both in the Newtonian regime as well as on cosmological scales, we need to state our assumptions for both regimes. In the rest of this paper we will proceed with these assumptions to analyze the effect on cosmological parameter extraction, and in particular on the inference of the Hubble constant.

We can summarize our assumptions of our hypothesis by the following combination of statements:

\begin{itemize}
\item The effective gravitational law in the Newtonian regime is an inverse square law with a coupling constant in the Newtonian regime for $t_0>t>t_T$ given by $G_N$. This is the constant measured in present-day laboratory tests of the inverse-square law and has a value $G_N = 6.67\times 10^{-11}$~N~m$^2$/kg$^2$. For notational simplicity, we now simply drop the subscript and refer to this constant as $G$.
\item The gravitational force law in the Newtonian regime for $t<t_T$ is still of the $1/r^2$ form but with effective coupling $G + \Delta G$. The Newtonian regime is assumed to be valid on stellar physics scales relevant for Cepheids and Type Ia supernovae.
\item The constants $\Delta G$ and $t_T$ are parameters of our model. One can alternatively use a transition look-back distance $d_T$ or transition redshift $z_T$ instead of the parameter $t_T$. We will consider a range of possible values for $\Delta G$ such that $0<\Delta G/G < 10\%$. The transition time $t_T$ is assumed to be close enough in our past such that at least some of the observed calibrator galaxies which host Cepheids lie at redshifts beyond $z_T$, and thus experience a different effective gravitational coupling. This constrains the range of transition redshifts to values $z_T\lesssim 0.02$.
\item The leading-order metric is described by a flat FRW universe with a $\Lambda$CDM type matter-energy content, with a modified law for cosmological evolution of the scale factor $a(t)$ given by,
    \begin{eqnarray}
    \label{eq:FRWGtransition}
\left ( \frac{\dot{a}}{a} \right ) ^2 =
\begin{cases}
 \frac{8 \pi G \rho_c}{3} \left(\frac{\Omega_m}{a^3}+ \frac{\Omega_r}{a^4}+  \Omega_\Lambda \right), \textrm{ for } t> t_T,\\
 \frac{8 \pi (G+\Delta G) \rho_c}{3} \left(\frac{\Omega_m}{a^3} +\frac{\Omega_r}{a^4} + \Omega_\Lambda \right), \textrm{ for } t< t_T.
 \end{cases}
\end{eqnarray}
Here $\Omega_m$, $\Omega_r$ and $\Omega_\Lambda$ are the usual present-day matter, radiation, and vacuum energy density fractions, and since we are assuming a flat FRW universe, we have $\Omega_r+\Omega_m+\Omega_\Lambda = 1$. Note that the constants $G$ and $G + \Delta G$ in this equation are the same as the ones that show up in the Newtonian regime.

\item The modified value of the coupling $G + \Delta G$ also determines the gravitational driving of the density perturbations in the early universe.
\item If we were to fit these assumptions into the framework of scalar-tensor theory, then one would in principle have to study the cosmological implications of the new scalar field. We assume that the scalar field responsible for the effective change in $G$ has no other significant cosmological effect --- in particular we assume that there is no contribution to the cosmological expansion history through its energy density, and also that there are negligible spatial inhomogeneities in the value of the scalar field, and hence in the value of the effective gravitational constant, throughout the universe.
\end{itemize}

In appendix~\ref{sec:scalartensor}, we discuss a self-consistent covariant candidate framework which generalizes Einstein's general relativity in which such a transition can occur --- namely scalar-tensor theory. A complete specification of how a $G$-transition can be arranged in this theory is beyond the scope of this work, but we show in principle the requirements necessary to build a model within the framework of scalar-tensor theory that can realize a late-time $G$-transition. Assuming that these requirements can be satisfied, these would then lead to our working assumptions.

\section{Inference of the Hubble constant and other cosmological parameters in the presence of a $G$-transition}
\label{sec:CMBinference}
What are the implications of a cosmological $G$-transition for cosmological parameter extraction, and in particular the extraction of the Hubble constant?

\vspace{3mm}
First we note that in the $G$-transition cosmology, given the assumptions laid out in sec.~\ref{sec:assumptions}, it makes sense to define not one, but rather two Hubble constants, $H_{0\alpha}$ and $H_{0\beta}$, as follows,
\begin{eqnarray}
\label{eq:H0A}
{H^2_{0\alpha}} &=&  \frac{8 \pi G \rho_c}{3}  , \\
{H^2_{0\beta}}   &=& \frac{8 \pi G\rho_c}{3}\left ( 1 + \frac{\Delta G}{G} \right  ).
\label{eq:H0B}
\end{eqnarray}

In the standard cosmology, one has only a single Hubble constant $H_0$ that can be fitted for from cosmological data sets. The parameter $\rho_c$ is then a dependent parameter that is extracted once the value of $H_0$ is known, assuming that the gravitational constant is simply $G=G_N$. In the $G$-transition scenario, the critical density $\rho_c$ and $\Delta G$ are both to be regarded as dependent parameters that can be determined from $H_{0\alpha}$ and $H_{0\beta}$.

Furthermore, when considering fits of cosmological parameters to the data, we will find it useful to separate our discussion into two categories depending on the redshift of the cosmological sources. We classify a high redshift data-set as one obtained from sources at redshifts $z \gg z_T$. This would include the CMB, baryon acoustic-oscillation data (BAO), large-scale structure (LSS) data etc. We classify a low redshift data-set as one which contains sources at $z \sim z_T$. This would include things like type Ia SNe, TRGBs etc.

Ideally, for either kind of data-set, one would like to reanalyze the full data and fit for the parameters of the $G$-transition cosmology. We will argue below that all the cosmological data sets can only be used to infer the value of $H_{0\beta}$, and are practically insensitive to the value of $H_{0\alpha}$. Both the CMB data and the low redshift data need to be reanalyzed to correctly infer the corresponding values of $H_{0\beta}$ from each data set. However, we shall argue that for CMB data one may simply recast the existing analyses of the standard cosmology to infer the parameters of the $G$-transition cosmology. On the other hand, for low redshift data-sets, we will argue that the recasting is not straight-forward and one needs to (at least partially) reanalyze the data to obtain an inference of the cosmological parameters.

\vspace{5mm}
Let us now justify these claims.

The key to connecting cosmological data to model parameters of either the $G$-transition cosmology or the standard cosmology is to first obtain the comoving distance to redshift relation $r(z)$. Once this is known, one can derive the luminosity $d_L(z) = (1+z) r(z)$ or angular diameter $d_A(z) = r(z)/(1+z)$ distances to a source at redshift $z$.

The comoving distance $r(z)$ to a source in any cosmology is given by,
\begin{eqnarray}
r(z) &=& c \int_t^{t_0} \frac{dt^\prime}{a(t^\prime)}= c \int_0^{z} \frac{dz^\prime}{H(z^\prime)}.
\end{eqnarray}

\vspace{3mm}
\textbf{Standard cosmology:} In the standard cosmology one can further use the FRW equations for the scale-factor evolution to obtain,
\begin{eqnarray}
\label{eq:rzLCDM}
r(z) &=& \frac{c}{H_0}\left[  \int_{0}^{z} f(z^\prime) dz^\prime\right ],
\end{eqnarray}
where,
\begin{equation}
\label{eq:fz}
 f(z) = \frac{1}{\left(\Omega_r (1+z)^4+ \Omega_m (1+z)^3+   \Omega_\Lambda \right)^{1/2}}.
\end{equation}
Note that since we are assuming a flat FRW universe where $\Omega_r + \Omega_m +   \Omega_\Lambda=1$, this implies $f(z=0)=1$.

\vspace{3mm}
\textbf{$\mathbf{G}$-transition cosmology:} In the $G$-transition cosmology, with the definitions of the effective Hubble constants above, we can rewrite the modified FRW equation for scale-factor evolution (eq.~\ref{eq:FRWGtransition}) in terms of redshift $z$, where $a = \frac{1}{1+z}$ as,
\begin{eqnarray}
\left ( \frac{\dot{a}}{a} \right ) ^2 = \left ( \frac{\dot{z}}{1 +z} \right ) ^2 =
\begin{cases}
{H^2_{0\alpha}} \left(\Omega_r (1+z)^4+ \Omega_m (1+z)^3+   \Omega_\Lambda \right) \textrm{ for } t> t_T, \\
{H^2_{0\beta}}  \left(\Omega_r (1+z)^4+\Omega_m (1+z)^3+   \Omega_\Lambda \right) \textrm{ for } t< t_T.  \\
\end{cases}
\end{eqnarray}

Since we have assumed a sudden $G$-transition, $dz/dt$ is discontinuous at $t_T$, but $z(t)$ itself is continuous. Thus, the above equation allows us to relate the transition time $t_T$ to a transition redshift $z_T = z(t_T)$. Now depending on whether we consider a source at $z>z_T$ or $z<z_T$, we have the following expressions for $r(z)$,
\begin{eqnarray}
\label{eq:rzlowGtrans}
\left . r(z)\right |_{z<z_T} &=& \frac{c}{H_{0\alpha}}\left[ \int_{0}^{z} f(z^\prime) dz^\prime \right ].
\end{eqnarray}
and
\begin{eqnarray}
\left . r(z)\right |_{z>z_T}  &=& c\left[ \frac{1}{H_{0\beta}} \int_{z_T}^{z} f(z^\prime) dz^\prime
    +  \frac{1}{H_{0\alpha}} \int_{0}^{z_T} f(z^\prime) dz^\prime \right ], \\
    &=& c\left[ \frac{1}{H_{0\beta}} \int_{0}^{z}  f(z^\prime) dz^\prime
    +  \left ( \frac{1}{H_{0\beta}} - \frac{1}{H_{0\alpha}}  \right ) \int_{0}^{z_T}  f(z^\prime) dz^\prime \right ],\\
    &=& \frac{c}{H_{0\beta}}\left[  \int_{0}^{z} f(z^\prime) dz^\prime  + k \right ].
    \label{eq:rzhighGtrans}
\end{eqnarray}
where in the second line we have made a suggestive reorganization of the terms by adding and subtracting an integral from $0$ to $z_T$, and in the last line we have defined the redshift-independent constant $k$ as,
\begin{eqnarray}
\label{eq:kdef}
k &\equiv& \left [ \frac{H_{0\beta}}{H_{0\alpha}} - 1 \right ] \int_{0}^{z_T} f(z^\prime) dz^\prime  , \\
&=&  \left [ \left ( 1 + \frac{\Delta G}{G_N} \right  )^{1/2} - 1 \right ] \int_{0}^{z_T} f(z^\prime) dz^\prime ,
\end{eqnarray}
which depends on $\Delta G$ and $z_T$ (and also the density fractions).
Note that since we are interested in transitions with $|\Delta G/G |< 10\%$ and $z_T < 0.02$ by assumption, we can approximate the constant $k$ as,
\begin{equation}
k \equiv  \left ( \frac{1}{2}\frac{\Delta G}{G_N} \right ) z_T,
\end{equation}
where we have set $f(z)\approx 1$ at very low redshifts, since in the integrand on the right-hand side of eq.~\ref{eq:kdef}, $z^\prime <z_T \ll 1$. Hence, at leading order, the constant $k$ is independent of the cosmological density fractions.

\vspace{3mm}
\textbf{Effect on cosmological parameter inference from the CMB and other high redshift probes}
For all redshifts relevant for high-redshift cosmological data-sets, the relevant $r(z)$ relation in the $G$-transition cosmology is given by eq.~\ref{eq:rzhighGtrans}. By comparing this $r(z)$ relation to that of the standard cosmology (eq.~\ref{eq:rzLCDM}), we can see that in the cosmology with a $G$-transition for $z > z_T$, the formula for $r(z)$ is almost identical to that of the standard cosmology except for,
\begin{itemize}
\item a replacement of $H_0$ by $H_{0\beta}$.
\item the inclusion of the constant $k$.
\end{itemize}
The $k$ dependent effects are highly suppressed with both an $\mathcal{O}(\Delta G/G)$ and an $\mathcal{O}(z_T/z)$ suppression. Thus at large cosmological redshifts, the main effect on the $r(z)$ formula which causes it to differ from that of eq.~\ref{eq:rzLCDM} is simply the substitution $H_0 \rightarrow H_{0\beta}$.

In principle one needs to redo the fits to the CMB data in the presence of a $G$-transition to check the consistency of the fits and to extract the value of $H_{0\beta}$.
Previous studies of signatures of a $G$-transition on cosmological data with Planck 2018 CMB data combined with BAO data~\cite{Wang:2020bjk, Ballardini:2021evv, Sakr:2021nja} and  Big Bang Nucleosynthesis (BBN)~\cite{Alvey:2019ctk} have suggested that a change in the gravitational constant of at most around 5\% is allowed between the present day and in the early universe at the 2$\sigma$ level. However, care must be taken to interpret the results of these studies more generally because the constraints on $\Delta G$ depend on the nature of cosmological assumptions. For example, Ballardini et al~\cite{Ballardini:2021evv} considered various realizations of a scalar-tensor where the gravitational constant on cosmological scales at $z=0$ can be different from the Newtonian gravitational constant. For CMB observations, such a change is nearly equivalent to studying a change like ours from $G$ to $G$ + $\Delta G$ at $z\simeq z_T$, since $z_T$ is much smaller than the redshift of recombination. However, the models of ref.~\cite{Ballardini:2021evv} also include additional effects such as a contribution to the number of effective neutrinos from the scalar degree of freedom, and also a time varying (decreasing) gravitational constant from the early universe to the present day. These latter effects are counter to our assumptions laid out in sec.~\ref{sec:assumptions}.

Instead of refitting the full CMB data to the $G$-transition cosmology, we will instead give arguments for how the corrections to the value of $H_0 = 67.66 \, \pm  \, 0.42$~\Hunit (68\% C.L.)~\cite{p2020} extracted by Planck can be computed and interpreted as the would-be fit value of $H_{0\beta}$, were we to redo the fits. The argument is as follows below.

In the standard cosmology, the Hubble constant $H_0$ can be extracted from observation of the angular size of the first peak in the CMB $(\theta_*)$. This angular size is theoretically given by the ratio of the physical sound horizon size at the surface of last scattering $r_s(z_*)$, to the angular diameter distance to this surface $d_A(z_*)$, i.e. $\theta_* = r_s(z_*)/d_A(z_*)$ where $z_* \simeq 1100$ is the redshift of the surface of last scattering from which the CMB is emitted.

The physical sound horizon $r_s(z_*)$ is given by,
\begin{equation}
r_s(z_*) = \frac{1}{1+z_*} \int_{z_*}^\infty \frac{dz}{H(z)} c_s(z),
\end{equation}
where $H(z)$ is the Hubble rate, and $c_s(z) = \sqrt{\frac{1}{3(1+R)}}$ is the sound speed, and the integral receives contributions from all redshifts beyond the last scattering surface. Here, $R= \frac{3}{4}\frac{ \rho_b}{\rho_\gamma}$ depends on the baryon-to-photon density. The sound horizon $r_s(z_*)$ can be well determined from knowledge of the CMB temperature and the ``potential envelope'' (which determines $\omega_m=\Omega_m h^2 $)~\cite{Knox:2019rjx}, without any reference to the Hubble constant. See appendix~\ref{sec:appendixB} for a calculation of $r_s(z_*)$.

The angular diameter distance to the surface of last scattering is given by,
\begin{equation}
\label{eq:dAzstar1}
d_A(z_*) = \frac{1}{1+z_*} r(z_*)
\end{equation}
where $r(z_*)$ is the comoving distance  to the last scattering surface and is given by eq.~\ref{eq:rzLCDM}. The dependence on the Hubble constant arises through the dependence of $r(z_*)$ on $H_0$, but care should be taken to keep track of the independent parameters of the fit which are usually taken to be the Hubble-weighted matter and baryon density fractions, $\omega_m = \Omega_m h^2$ and $\omega_b = \Omega_b h^2$ rather than $\Omega_m$ and $\Omega_b$. With this choice of parameterization we have,
\begin{equation}
\label{eq:dAzstar2}
d_A(z_*) \simeq \frac{1}{H_{100}}\frac{1}{1+z_*}  \bigints_{0}^{z_*} \frac{1}{\left[ \omega_m (1+z)^3+   \left(\frac{H^2_0}{H^2_{100}}- \omega_m \right) \right]^{1/2}} dz,
\end{equation}
where we have defined $H_{100} = 100$~\Hunit, and we have neglected the radiation contribution to the angular diameter distance (which is a good approximation to within 0.5\%). This integral needs to be numerically computed to determine the relationship between $d_A(z_*)$ and $H_0$ (assuming that we know the value of $\omega_m$). We can then match the angular diameter distance to the one predicted by the measured angular size of the first peak and the size of the sound horizon, to determine the Hubble constant.

In the $G$-transition cosmology, the changed value of the gravitational constant in the early universe has possible effects on both $d_A(z_*)$ as well as on $r_s(z_*)$.

Let us first consider the effect of a $G$-transition is on the prediction of $r_s(z_*)$. One has to carefully keep track of the independent parameters of the CMB fit to see where (and why) the value of $G$ shows up in the theoretical calculation of $r_s(z_*)$. The Hubble constant $H_0$, and the Hubble-weighted matter and baryon density fractions, $\omega_m$ and $\omega_b$ are usually regarded as independent fit parameters and the critical density $\rho_c$ is a derived parameter. With this choice of parameterization, the value of $G$ shows up when we replace $\rho_c$ with the independent parameters.

In appendix~\ref{sec:appendixB}, we  show that a change in $G$ (while holding the independent parameters fixed) can affect $r_s(z_*)$ through both $H(z)$ in the pre-recombination era as well as $c_s(z)$. We perform a numerical estimate of the effect of a change in $G$ on $r_s(z_*)$ while keeping the parameters $\omega_m $ and $\omega_b$ fixed to the Planck values. We find that, $\frac{\Delta r_s(z_*)}{r_s(z_*)} \simeq -0.16 \frac{\Delta G}{G}$.

Next let us consider the effect of a $G$-transition on the prediction of $d_A(z_*)$. The expression for the angular diameter distance in eq.~\ref{eq:dAzstar1} must be modified by expressing $r(z_*)$ using eq.~\ref{eq:rzhighGtrans}, which is the appropriate expression for the $G$-transition cosmology. The effect of the $k$ dependent correction term on the angular diameter distance is suppressed by at least $O\left(\frac{\Delta G}{G} \frac{z_T}{z_*} \right) \simeq 0.1 \times \frac{0.01}{1100} \simeq 10^{-6}$ and can safely be ignored when fitting for cosmological parameters given the precision of the Planck collaboration's parameter extraction~\cite{p2020}. Thus, the main change in the expression for $d_A(z_*)$ is a replacement of $H_{0}$ by $H_{0\beta}$ in eq.~\ref{eq:dAzstar2}.
Thus, in the case of the $G$-transition cosmology, it is the constant $H_{0\beta}$ that can be extracted from the knowledge of the sound horizon size $r_s(z_*)$ and the angular size of the first peak $(\theta_*)$.

In appendix~\ref{sec:appendixB}, we also show that the change in inferred value of $H_{0\beta}$ due to the change in the sound horizon in the presence of a $G$-transition is given by,
\begin{equation}
\label{eq:deltaH}
\frac{\Delta H_0}{H_0} \simeq -\frac{1}{0.19}\frac{\Delta r_s(z_*)}{r_s(z_*)} \simeq +0.83 \frac{\Delta G}{G},
\end{equation}
where $\Delta H_0 = H_{0\beta} - H_0$ and $H_0$ is the value inferred from the CMB assuming the standard cosmology.
To obtain this relationship we have assumed that the fitted value of cosmological parameters other than $H_{0\beta}$ are unchanged from the standard cosmology. With this assumption, we see that for a positive $\Delta G$, $H_{0\beta}$ is larger than the Planck value, thus potentially ameliorating the Hubble tension.

The analysis above is only indicative, as it does not involve a full CMB fit. Upon performing such a fit, it is possible that the values of the other cosmological parameters may change and this will modify the inference of $H_{0\beta}$. For the moment we will ignore this putative change in the value of $H_{0\beta}$ and we will simply assume that the value of $H_{0\beta}$ that would be extracted from refitting the CMB data assuming a $G$-transition cosmology would be the same as the Planck inferred value of $H_0$, i.e. we will ignore the change $\Delta H_0$. We will later comment on the implication of an inferred value of $H_{0\beta}$ that is larger than the Planck value.

\vspace{5mm}
\textbf{Effect on cosmological parameter inference from type Ia SNe and low redshift probes}

The arguments for the replacement of $H_0$ by $H_{0\beta}$ while keeping other cosmological parameters fixed relies on the similar forms of the $r(z)$ relations in the $G$-transition and standard cosmology. This similarity is only approximately valid when we can ignore the effect of the constant $k$ in eq.~\ref{eq:rzhighGtrans}. At high redshifts $k$ is highly suppressed due to the suppression factor $\frac{z_T}{z}$. However for low redshift probes such as type~Ia~SNe, where some SNe are at $z\sim \mathcal{O}(1) \times z_T$ this raises a natural concern about whether similar arguments can be applied to the inference of cosmological parameters and the Hubble constant.

We will explicitly derive the luminosity-distance redshift relation $d_L(z)$ in the next section (specifically in sec.~\ref{sec:SNedlzrelation}) and we will use this to argue that the $k$ dependent corrections to the luminosity distances are potentially of importance if $z_T$ is near the upper end of our range, i.e. $z_T \sim 0.02$. However, given the redshifts of these SNe, and the current percent level of precision of determination of the Hubble constant, if $z_T$ is smaller than this maximum value by a factor of a few (say 3-4 times smaller), then the $k$ dependent corrections are not very important. However, in either of these cases, it is only the value of $H_{0\beta}$ that can be extracted from these low redshift probes.

\vspace{3mm}
If both the high and low redshift data are probing a single constant $H_{0\beta}$, even in the $G$-transition cosmology, then one may wonder how this could possibly resolve the Hubble tension since one returns to the problem raised in the introduction --- which is that both these values appear to be discrepant. As we shall discuss in the next section, the reason for the mismatch of the two inferred values of the Hubble constant is that the behaviour of distant SNe (beyond $z_T$) is different from those of nearby SNe since they experience different gravitational coupling constants (and also potentially a modified $d_L(z)$ relation). This difference in behavior needs to be properly accounted for when fitting the low redshift data. We shall argue in the next section that correctly accounting for this difference can resolve the Hubble tension.

\vspace{5mm}
To summarize this section, we have seen that the Hubble constant being probed in the $G$-transition cosmology is the parameter $H_{0\beta}$ and the value of this as inferred from CMB data is nearly the same as (or even slightly larger than) the value of $H_0$ assuming the standard cosmology. This motivates us to identify $H_{0\beta}$ in the $G$-transition cosmology as the closest analogue of $H_0$ in the standard cosmology. The Hubble tension is a discrepancy between the value of $H_0$ as measured from the CMB and low redshift SNe assuming the standard cosmology. This discrepancy will be resolved by appropriately taking into account the effects of a $G$-transition while interpreting the low redshift measurements.

\section{The distance ladder and the effect of a $G$- transition}\label{sec:intuition}
In the standard cosmology,  the expansion rate can be approximately described in the local universe by a linear relation $v = H_0 \, r$. Here, $v$ is the recession velocity of a galaxy located at a distance $r$. This is commonly known as the Hubble - Lema{\^\i}tre law~\cite{Lematre1927, Hubble1929} or Hubble law in short. The value of $H_0$ can be determined by finding the recession velocities and distances to distant objects and fitting to the linear relationship expected from the Hubble law. Velocities can be measured by redshifts of characteristic spectral lines and distances can be measured by constructing a distance ladder through some standardized astrophysical objects. In order for the Hubble law to hold, distances (or redshifts) need to be large enough so that the recession velocity is larger than the peculiar motions due to local gravitational flows. Typically this condition is satisfied for galaxies at redshifts $z \gtrsim 0.01$, for which the recession velocity is primarily due to cosmic expansion. Such galaxies are said to belong to the \enquote{Hubble flow}. Thus, $H_0$ can be determined by fitting a distance-redshift relation to galaxies in the Hubble flow, for which the Hubble law can be restated in the more useful form,
\begin{equation}
\label{eq:Hubblelawstd}
d_L(z) = \frac{c}{H_0}z,
\end{equation}
where $d_L$ is the luminosity distance to a galaxy at redshift $z$. This law is only approximate for low redshifts and the right-hand side should really be regarded as the first term of a Taylor series expansion in $z$.

A transition in the gravitational coupling constant will have two distinct types of effects on the interpretation of the distance ladder and hence on the inference of the Hubble constant. First, such a transition will alter the astrophysics of standard objects used in constructing the distance ladder, and second it will also alter the luminosity-distance redshift relation, and potentially even the Hubble law itself. In the rest of this section, we will describe how the standard cosmic distance ladder is constructed using observations of Cepheid variables and type Ia  SNe. We will then discuss how distance measurements, and consequently the inferred value of the Hubble constant, would be affected by a sudden $G$-transition. Our treatment in this section is purely analytic so as to clearly delineate the effect of a $G$-transition on the inference of the Hubble constant.

\subsection{The standard distance ladder}

The Supernovae and $H_0$ for dark energy Equation of State (SH0ES) collaboration has claimed the most precise local measured value of $H_0$~\cite{R16,R19,R21}. The SH0ES team primarily studied luminous type Ia SNe in the Hubble flow which are well known to be ``standardizable candles'' \cite{Phillips93, Carroll2001} . The following discussion will describe the strategy for the SH0ES analysis and how they establish a distance ladder to calibrate type Ia SNe.

The progenitor for an SN Ia explosion is believed to be accretion or merger of a white dwarf in a binary system~\cite{SN-CO}. When the white dwarf nears the Chandrashekhar mass, it undergoes runaway nuclear fusion that unbinds the star in a catastrophic explosion. Because of the fixed critical mass of the progenitor, type Ia SNe are expected to have a standard luminosity, i.e. they are expected to be standard candles \cite{Branch92}. If we know this standard expected luminosity, then we can combine this with the flux measurement from observed type Ia SNe to measure distances to galaxies in the Hubble flow which host such SNe.

Observed type Ia SNe explosions have variable peak luminosities and therefore are not truly standard candles \cite{Phillips93, Carroll2001}. However, the peak luminosities are tightly (positively) correlated with the decay time of the light curve for such SNe \cite{Phillips93, Carroll2001}. Observations of Hubble flow SNe at a given redshift indicate that ``stretching'' the light curves to agree with the shape of a template light curve yields nearly identical light curves~\cite{Tripp1998, Betoule2014}. This allows us to standardize the light curves to a template at a given redshift. Moreover this standardization of the light curves to a given template (with a suitable redshift correction to the apparent magnitude) works at all redshifts in the Hubble flow, indicating very little evolution~\cite{Scolnic2018} with redshift of the intrinsic standardized type Ia SNe light curve (after taking into account various other corrections like dust extinction, coherent flows in the local universe etc.).

If we knew the intrinsic peak luminosity of the standardized template, we could use this to infer the distance to type Ia SNe. Thus, type Ia SNe are referred to as ``standardizable candles''. In order to use type Ia SNe to measure the Hubble constant, one needs to first calibrate the standardized peak luminosity of nearby SNe Ia. The peak luminosity can then be inferred from a combination of knowledge of the flux and distance to a type Ia SN in a nearby galaxy. Distances within our galaxy and nearby galaxies can be directly determined with high precision through either trigonometric parallaxes \cite{Lindegren2021, R21}, Detached Eclipsing Binaries (DEBs) \cite{Paczynski97}, or water MASERs \cite{Herrnstein99}. However, no SN Ia explosion has been observed to which such a direct distance measurement is available.

Thus, the standard Type Ia SNe luminosity needs to be calibrated with an intermediary. The SH0ES collaboration uses classical Cepheid variables as the intermediary. Cepheids are pulsating stars, where the pulsations are driven by the Eddington valve or $\kappa$-mechanism \cite{edi18, book_edi}. Cepheid variables as discovered by Henrietta Leavitt have a well defined Period - Luminosity Relation (PLR) which allows them to be used as standard candles \cite{Leavitt1908, Leavitt1912}. Moreover, Cepheids are bright enough to be observable out to large extra-galactic distance scales with the Hubble Space Telescope \cite{Hoffmann2016,R16}. This makes it possible to find a sample of galaxies which a) host a SN Ia explosion, and b) contain a large number of Cepheid variables. The distances to these ``calibrator'' galaxies can be determined using the Cepheid PLR and then the SNe Ia luminosity can be derived. In order to use Cepheids to calibrate type Ia SNe, the standard Cepheid PLR needs to be first determined using observations of Cepheids in nearby ``anchor'' galaxies to which direct distance measurements are available.

Thus, the SH0ES analysis of type Ia SNe in the Hubble flow to measure the Hubble constant uses a distance ladder which involves the following three steps:
\begin{itemize}
    \item \textbf{Anchor step}: This step involves calibrating the standard Cepheid PLR with the help of geometric distances. Cepheids in the MilkyWay (MW) and nearby galaxies like the Large Magellanic Cloud (LMC) and NGC4258 are used for this purpose. For Cepheids in the MW, LMC, and NGC4258, SH0ES uses trigonometric parallax based distances \cite{Lindegren2021, R21}, DEB based distances \cite{P19, R19}, and water MASER based distances \cite{H13, R16}, respectively. Knowledge of these distances along with the measured fluxes and periods of the Cepheids yields the PLR. The anchor objects have distances up to approximately 7 Mpc.
    \item \textbf{Calibrator step}: This involves calibrating the SNe Ia luminosity with the Cepheid PLR. A set of $37$ calibrator galaxies which have had SNe Ia explosions and also contain Cepheid variables are used for this purpose \cite{R22}. The PLR derived from anchors, along with the measured Cepheid periodicity is used to infer distances to these calibrator galaxies. These distances combined with the corrected peak apparent magnitudes of type Ia SNe yield their intrinsic standardized peak-luminosity. The calibrator galaxies range in distances from approximately 7~Mpc to 80~Mpc.
    \item \textbf{Hubble flow step}: Finally, the standardized luminosity of SNe Ia are used to infer the distances to Hubble flow SNe. By measuring the redshift of the host galaxies, SH0ES finds a distance-redshift relation for several hundred SNe Ia in the Hubble flow and they use this to determine the value of $H_0$. It includes SNe Ia ranging from redshift $z=0.023$ to $z= 0.15$ (corresponding to distances $>$ 80 Mpc).

\end{itemize}

In practice, the SH0ES team performs a simultaneous fit to data for the anchors, calibrators, and Hubble flow objects. In the next sub-section, we will explain how a $G$-transition will affect this distance ladder, and alter the inference of the Hubble constant.

\subsection{Effect of a $G$-transition on the distance ladder}

We have defined a gravitational constant ($G$) transition as a sudden change in the value of $G$ at some cosmic epoch. The value of $G$ at the present epoch is taken to be $G_N= 6.67 \times 10^{-11}$~N-m/kg$^2$ as measured in terrestrial experiments~\cite{10.1093/nsr/nwaa165}. Instead of referring to a transition time $t_T$ or redshift $z_T$ before which the gravitational constant was larger by an amount $\Delta G$, we can equivalently parameterize the transition to occur at some transition distance $d_T$\footnote{The transition distance $d_T$ here refers to a luminosity distance, but at the very low redshifts at which we are studying a $G$-transition, the distinction between comoving distance and luminosity distance measure of $d_T$ is practically irrelevant.}. Here, $d_T$ and $\Delta G$ are the extra parameters of our model. In the rest of this article whenever we refer to objects which lie to the left or to the right of the transition, this should be taken to mean objects at $d < d_T$ or $d > d_T$, respectively.

If the $G$-transition occurs sufficiently late in our cosmological history, i.e. for sufficiently low $d_T$ below a few 100~Mpc, it would directly affect the properties of the objects that constitute the distance ladder beyond $d_T$, and hence alter the inferred value of $H_0$. Depending on the precise value of $d_T$, it would modify the standardized SNe Ia peak-luminosity for some/all Hubble flow supernovae, or for sufficiently low $d_T$, it could possibly even alter the Cepheid PLRs.

A transition at $d_T >80$~Mpc would affect the standardization of type Ia SNe light curves in the Hubble flow and might be in conflict with observations which have indicated no evolution in SNe~Ia light curve properties. A transition distance $d_T = 80$~Mpc  at the boundary between calibrators and the Hubble flow SNe was proposed in ref.~\cite{Marra:2021fvf} in an attempt to solve the Hubble tension. However, in order to alleviate the tension, this study assumed a peak-luminosity -- Chandrashekhar mass relation $L\propto M_c$, an assumption which is in contradiction with the results of the semi-analytic model of ref.~\cite{Wright2018} which indicates an inverse relationship between $L$ and $M_c$.

In the present work, we focus on a transition that occurs within the set of calibrator galaxies which lie at distances between $\sim 7$ - $80$~Mpc (SH0ES refers to this as the calibrator rung of the distance ladder) in such a way that some calibrators lie beyond the distance $d_T$. It will become clear later when we discuss the distance-redshift relation in sec.~\ref{sec:SNedlzrelation} that this assumption translates to $z_T \lesssim 0.02$, which motivates the range of values of $z_T$ that we decided to restrict to\footnote{Strictly speaking the relationship between $d_T$ and $z_T$ depends upon the precise value of the Hubble constant, but at such low redshifts, and given the level of accuracy we are interested in, the precise value is not so important.}. Such a transition would lead to different properties of the nearby Cepheids and SNe, as compared to those beyond $d_T$. Cepheids and SNe Ia can still be used as standard candles, but their standard calibrations will be different before and after the epoch of $G$-transition. Thus, if the hypothesis of a $G$-transition is correct, not accounting for the change in properties of the distant Cepheids and SNe Ia would lead to an incorrect inference of their distances and hence an incorrect inference of the Hubble constant. This could potentially explain the discrepancy between the local and distant universe measurements of $H_0$.

In the next few sub-sections we will explain the effect of a $G$-transition on the distance ladder. We will first explain the effect of a $G$-transition on the Cepheid PLR and SNe Ia standardized peak luminosity. We will then explain, for each of these objects, how not taking into account these changes would lead to an incorrect inference of the distances to their host galaxies. Next, we will discuss the effect of a $G$-transition on the luminosity-distance redshift relation which is relevant for Hubble flow SNe. We will then put these two effects together to explain how the incorrectly inferred distances would lead to an incorrectly inferred Hubble constant.

\subsubsection{The Cepheid PLR and a $G$-transition}
\label{sec:cep_PLR_G}
Cepheids are variable stars, which populate the upper region of the instability strip in the optical color-magnitude parameter space~\cite{Alfred96}. The instability strip refers to a region of the Hertzsprung–Russell (HR) diagram where a star suffer instabilities causing it to pulsate in size and in luminosity. Cepheid variables act as standard candles for extra-galactic distance determination because of their tight period-luminosity relation~\cite{Leavitt1908, Leavitt1912}. They are also luminous enough in order to be observable out to large extra-galactic distance scales up to $80$ Mpc with the Hubble Space Telescope (HST)~\cite{R16, R22}. The Cepheid PLR more generally can be expressed as a period-luminosity-color relation (PLCR) as follows~\cite{book_edi},
\begin{equation} \label{eq:basic_PLCR}
    \log(L) = a \log\left(\frac{P \: \rm{(days)}}{10 \: \rm{days}}\right) + b \log\left(T_{\textrm{eff}}\right) + c
\end{equation}
here $L$ is the mean luminosity\footnote{Since the luminosity periodically changes with time, the PLR is expressed using the mean luminosity.}, $P$ is the pulsation period, and $T_{\textrm{eff}}$ is the effective surface temperature of the Cepheid which can be replaced by observable colour. Here, $a$, $b$, and $c$ are coefficients that determine the PLCR. When Cepheids are observed in a given wavelength band (colour or effective temperature is fixed), the $3$-dimensional PLCR gets projected on to the $2$-dimensional period-luminosity plane and this leads to the PLR~\cite{book_edi},
\begin{equation} \label{eq:basic_PLR}
    \log(L) = \alpha \log\left(\frac{P \: \rm{(days)}}{10 \: \rm{days}}\right)  + \gamma.
\end{equation}
The coefficient $\alpha$ (which is positive) determines the slope of the PLR and $\gamma$ is the intercept. The coefficient $\gamma$ can also be thought of as the logarithm of the luminosity of a classic Cepheid variable with a period $P = 10$ days.

\vspace{3mm}
\textbf{Effect of a $G$-transition on the Cepheid PLR}

If the effective $G$ were different, this would change both the pulsation period, as well as the luminosity of a Cepheid. The resulting changes in the Cepheid period and luminosity would modify the Cepheid PLR.

The dynamics of Cepheid pulsations, and hence the pulsation period, are governed by the helium partial ionization zone which lies in the envelope of the star. On the other hand, the luminosity of the Cepheid is dictated by nuclear burning in the core~\cite{book_edi}. Thus, the change in the period and the change in mean luminosity can be analyzed independently to a good approximation.

Ritter~\cite{ritter} for the first time demonstrated that the pulsating period of a homogeneous sphere undergoing adiabatic radial pulsation varies with the mean surface density of the sphere as $P \propto \sqrt{R/g}$ where $R$ is the radius of the gaseous sphere and $g$ is the surface gravity. Later, many studies \cite{book_edi,martin, edi18, edi19} showed that this relationship is also valid for real stars. Heuristically, the pulsation period can be set proportional to the free-fall time of the Cepheid envelope, which scales as $P \propto 1/\sqrt{G \overline{\rho} }$ \cite{Sakstein2019}, where $\overline{\rho}$ is the mean density.

A detailed estimate of the scaling of $\overline{\rho}$ with $G$ would require modelling the physics of Cepheids in a modified $G$ environment. The density will be determined through the equilibrium dynamics of the envelope, which depends on a balance between pressure and gravity. The final scaling of $\overline{\rho}$ with $G$ that one would obtain would depend on factors such as the scaling of the opacity with density and the adiabatic index of the envelope. We will for simplicity assume that $\overline{\rho}$ is independent of $G$ (or equivalently exhibits a weak scaling with $G$). We will later comment on what would happen if $\overline{\rho} \propto G^m$, where $m\neq0$ is a scaling index.

With the assumption that the mean density of the envelope is unchanged by a change in $G$, this leads to a scaling relation $P \propto 1/\sqrt{G }$. Proceeding with this assumption we find that if the change in effective $G$ is $\Delta G$, the change in Cepheid period would be,
\begin{equation} \label{eq:P_change}
    \Delta \log (P) = - \frac{1}{2} \log \left(1 + \frac{\Delta G}{G}\right).
\end{equation}
In particular for a positive change $\Delta G$, the Cepheid period would decrease.

Now let us discuss the change in luminosity of a Cepheid due to a change in $G$. Cepheid variables burn a H-shell surrounding an inert He core (although, some amount of He core burning can take place)~\cite{book_edi}. For a fixed Cepheid mass, a slight increase in the effective gravitational constant would require more pressure support to maintain hydrostatic equilibrium. This pressure support can only be generated by increased nuclear burning in the core. The net result would therefore be an increase in luminosity.

Sakstein et al.~\cite{Sakstein2019} ran simulations with the MESA (Modules for Experiments in Stellar Astrophysics)~\citep{Paxton_2019} code, by modifying $G$ in the Cepheid cores. They obtained an expression for the change in luminosity at the blue edge of the instability strip which is of the form,
\begin{equation}
\label{eq:lum_change}
    \Delta \log L = B \, \log(1 + \Delta G /G),
\end{equation}
where the value of the coefficient $B$ depends on the mass of the Cepheid as well as which crossing of the instability strip is being considered\footnote{The crossing here refers to how many times the star crosses instability strip in the HR diagram during its evolution.}. In ref.~\cite{Sakstein2019} the authors tabulated values of $B$ as a function of the stellar mass and the instability strip crossing epoch. The typical values of $B$ that they obtained were between $3.46$ and $4.52$. Since $B$ is positive, this implies an increase in luminosity for an increase in the effective gravitational constant.

\begin{figure}
    \centering
    \includegraphics[width = 0.65\textwidth]{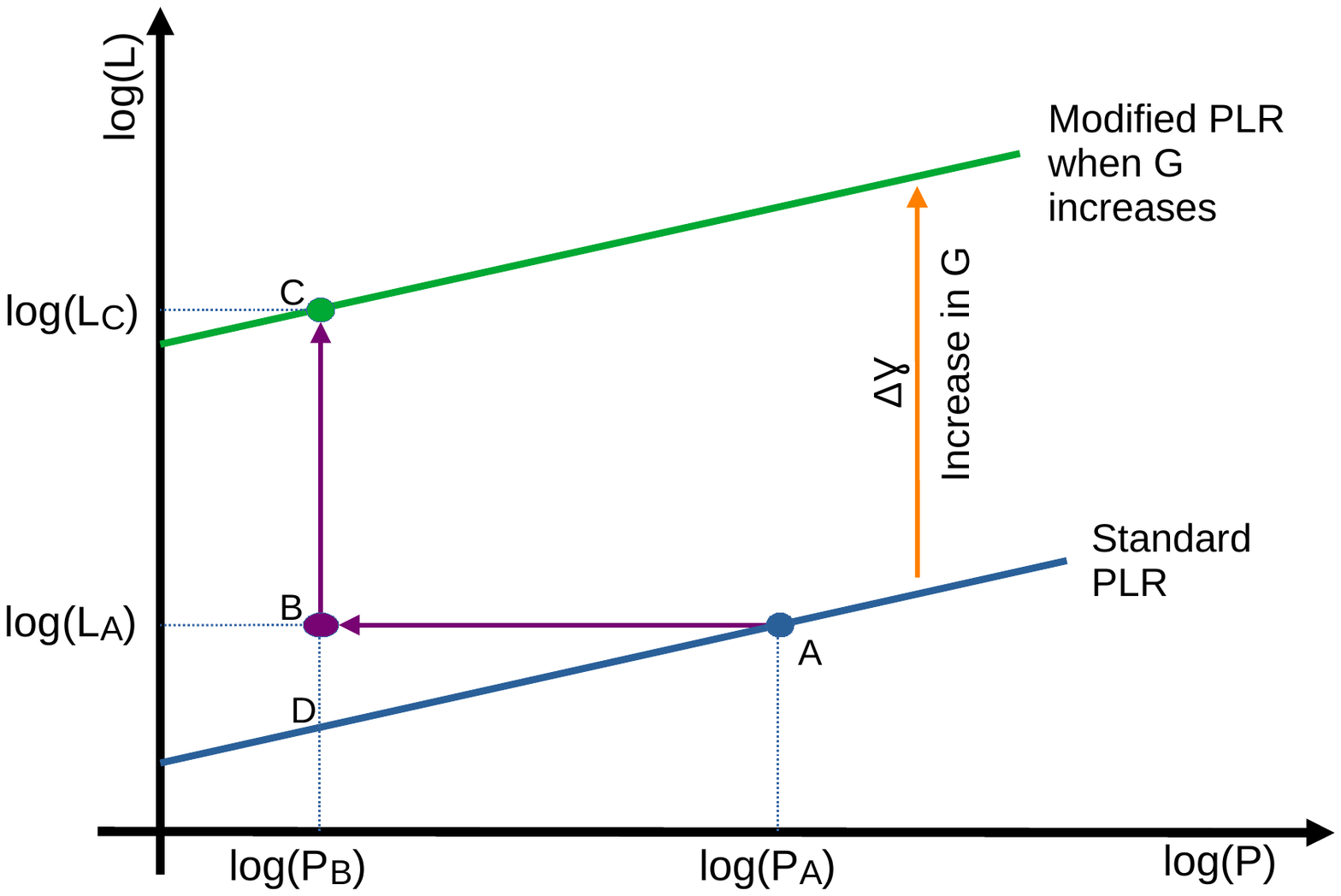}
    \caption{Illustration of how the Cepheid PLR would change if the effective $G$ were to increase. The original Cepheid PLR is shown in blue. Under an increase in $G$, a Cepheid at point A would have its period decrease and luminosity increase so that it would move to point $C$. By similar consideration for all other Cepheids, we would find that the effective PLR would be modified from the blue line to the green line, which has an identical slope to the original PLR but has an intercept difference $\Delta \gamma$. For Cepheids beyond the transition distance $d_T$, they would follow the green PLR. If we incorrectly used the low-distance blue PLR to infer their luminosities from their periods, we would underestimate the true log luminosity by an amount $\Delta \gamma$.}
    \label{fig:cartoon}
\end{figure}

The change in Cepheid PLR due to a change in $G$ can now be understood through a combination of the changes in the period and luminosity of a given Cepheid.
To illustrate this, we show a schematic diagram of the Cepheid PLR in fig.~\ref{fig:cartoon}. The blue line in the figure represents the standard Cepheid PLR when the gravitational constant is the standard $G$.

Now consider a specific Cepheid at point $A$ with luminosity $L_A$ and period $P_A$. What happens to the the period and luminosity for this Cepheid after a change in $G$? The period of the Cepheid would change to $P_B$, where  $\log(P_\textrm{B}) - \log(P_\textrm{A}) = - \frac{1}{2} \log \left(1 + \frac{\Delta G}{G} \right)$ and its luminosity would increase from $L_A$ to $L_C$, where $\log(L_\textrm{C}) - \log(L_\textrm{A}) = B \, \log \left (1 + \Delta G /G \right)$. This change is shown in the figure as a two step change with first the change in period only (point $B$), followed by a change in luminosity (point $C$). Thus, the Cepheid's position on the period-luminosity diagram would change to point $C$. Repeating this procedure for all Cepheids on the original PLR, we would obtain the modified PLR relation (shown as the green curve in the figure),
\begin{equation} \label{eq:basic_PLR_with_G}
    \log(L) = \alpha \log\left(\frac{P \: \rm{(days)}}{10 \: \rm{days}}\right)  + \gamma + \Delta \gamma,
\end{equation}
where the change in the PLR intercept $\Delta \gamma$  is
\begin{equation} \label{eq:PLR_intercept_G_1}
    \Delta \gamma = \left(\frac{\alpha}{2} +B \right) \log \left (1 + \frac{\Delta G}{ G} \right) .
\end{equation}
Thus, the net effect of a positive $(\Delta G > 0)$ $G$-transition is a modified Cepheid PLR with exactly the same slope as the original PLR but an increased intercept.

\vspace{3mm}

\textbf{Error in distance measurements due to assumption of a single PLR in the presence of a $G$-transition}

Now under our hypothesis of an effective gravitational constant $G$ at distances below $d_T$, and $G + \Delta G$ at distances larger than $d_T$, we can ask what incorrect inference would we make about the distances to Cepheids if we assumed a single PLR was valid at all distances?

Assuming that the calibration of Cepheids at lower distances yielded a PLR similar to the blue line in figure~\ref{fig:cartoon}, Cepheids beyond the transition distance $d_T$ would actually obey the modified (green) PLR. If we incorrectly used the blue PLR to infer the luminosity for a given observed pulsation period, we would
underestimate $\log L$ by the PLR intercept offset $\Delta \gamma$.

Thus, for $\Delta G$ positive we would underestimate the intrinsic luminosity of Cepheids beyond the transition distance. It is then easy to see that the error we would make on the inferred (luminosity) distance to far away Cepheids beyond $d_T$ would be,
\begin{equation}
\label{eq:cepheidcorr}
\log d_\textrm{inferred} - \log d_\textrm{true}= -\frac{1}{2}\Delta \gamma.
\end{equation}
Thus, for positive $\Delta G$, we would infer distances to far away Cepheids that are \textit{smaller} than what they truly are.

\subsubsection{Standardized SNe~Ia luminosity and the gravitational constant} \label{sec:SNeIaG}
SNe~Ia explosions are thought to occur in systems where a Carbon-Oxygen~(CO) white dwarf (WD) either merges with, or accretes mass from a binary companion. A WD is made of degenerate electron matter. For non-relativistic electrons in such stars, a WD has an inverse relationship between its mass and its radius. As a WD accretes matter, this would lead to further compression of the star, increasing its density and temperature. When the temperature reaches a critical threshold, which happens when the WD mass reaches the Chandrasekhar mass ($M_{\textrm{Ch}}$ $\approx$ 1.38 $M_\odot$)~\citep{SN-CO,CO1996,articlenickel}, rapid carbon detonation is triggered, leading to runaway nuclear fusion. This detonation can take place throughout the interior of the star since the interior of a WD is highly conducting  \citep{koester}. The runaway reaction destroys the star completely leaving behind no remnant and yielding an extremely luminous supernova with a total energy output near~$10^{51}$~erg over a few second burst \cite{woosley,sn251}. Most of this energy output of type Ia SNe is in the form of ejecta kinetic energy, with a sub-percent level of energy released into electro-magnetic radiation. This makes type Ia SNe some of the most luminous objects in the cosmos~\cite{Scalzo2014}. The power of a SN Ia luminosity in optical wavelengths is thought to dominantly arise from the decay chain of the Nickel-56 isotope produced in the explosion~\citep{articlenickel,Pinto}. It has been found that most luminous type~Ia SNe~(SN 1999aa and SN 2013aa) have anomalously high concentrations of the Ni-56 isotope~\citep{childress}.

Since the macroscopic conditions for all SNe~Ia progenitors are the same, one might naively expect that they should behave like standard candles with a fixed luminosity. Observed SNe~Ia have a light curve that increases rapidly over 10 - 20 days and then decays slowly over more than a month~\citep{parrent}.

Contrary to this naive expectation, observations of nearby SNe~Ia indicate that all type Ia SNe do not have a common peak luminosity. However, they do obey, to a very good approximation, a width-luminosity relation (WLR) which is the relation between SNe peak brightness and the time scale over which this peak brightness is achieved and then subsequently decays ~\cite{Phillips93}. This allows us to use SNe~Ia as~\textit{standardizable candles} by using their widths to infer the peak luminosity. In practice, the standardization is done by ``stretching'' the SN light curve to match a standard template. Given a particular amount of stretching, this allows one to define a correction factor to the observed apparent magnitude ~\cite{Betoule2014, Scolnic2018}.

Using SNe in calibrator galaxies to which distances are known through a calibrator (such as Cepheid variables), and combining this with the corrected apparent magnitude of these SNe then provides a standardized SN absolute magnitude which is assumed to be independent of distance/redshift.

Now, in order to measure the distances to Hubble flow SNe, we use their corrected apparent magnitudes along with the knowledge of the standardized absolute magnitude obtained from the SNe~calibrators to obtain the distance to a given Hubble flow SN host galaxy.

\vspace{3mm}

\textbf{Effect of a $G$-transition on the SNe~Ia standardized peak luminosity}

Type Ia SNe explosions are complicated to model because of the turbulent nature of the explosion and possible spontaneous transitions to detonation \citep{Pinto}. In principle, numerical models of SNe~Ia could be used to study the effect of a change in $G$ on the expected SNe~Ia standardized peak luminosity. However, here we make some simple assumptions to provide a simple analytic expression for the change in standardized peak luminosity due to a $G$-transition.

A first guess as to how the SNe Ia standard luminosity depends on $G$ is to assume that the peak luminosity scales in direct proportion to the Chandrasekhar mass~$M_\textrm{Ch}$ ~\cite{woosley, amendola,Gazt}. This mass is not very different from the Chandrashekhar limit $M_c \approx 1.44$~$M_\odot$ ~\cite{CO1996} (where relativistic degeneracy pressure is insufficient to protect the star against gravitational collapse), where $M_c \sim G^{-3/2}$ \cite{CO1996}. The inverse dependence of $M_c$ on $G$ can be easily understood. If $G$ is lower than the usual value, the gravitational pull per unit mass would become smaller and  therefore electron degeneracy pressure can counteract against gravitational pull produced by a larger mass just before the collapse happens. Thus, a star of higher mass can be supported against gravitational collapse, i.e. $M_c$ is higher for lower $G$. We will assume that $M_\textrm{Ch}$ also has the same scaling with $G$.

The above assumptions would therefore imply that the standardized SNe Ia luminosity $L \propto G^{-3/2}$, i.e. the luminosity decreases for an increase in $G$. However, as mentioned in the introduction, a semi-analytic model of SNe light curves by Wright and Li~\cite{Wright2018} suggests that the standardized SNe Ia luminosity might actually increase for larger values of $G$.

In order to provide an intuitive explanation for their results, we first explain a little bit of SNe~Ia physics. The luminous power of SNe~Ia is expected to arise dominantly from the decays of Nickel-56 which is produced in the explosion. The radiation from this decay must penetrate a dust cloud of ejecta around the supernova in order to escape. A plausible explanation for the variability in the luminosity of SNe Ia is the scatter in the amount of Ni-56 that is produced in the turbulent explosions. The width of the light curve on the other hand depends on the properties of the dust cloud such as its mass and opacity. Wright and Li constructed a semi-analytic model of SNe~Ia light curves and they argued that the tight observed WLR relation can be understood from a feedback effect of Ni-56 decays on the ionization of the ejecta and hence the opacity. This relates the total mass of Ni-56 to the opacity. They fixed this relationship so that the stretched light curves matched a standard template. The free parameters of their model that have the most dominant effect on the properties of the light curve are thus, the total mass of ejecta $M_\textrm{ej}$, and the total mass of nickel-56 produced $M_{\textrm{Ni}}$.

In this same work, the authors also examined the effect of a change in $G$ on the standardized SNe~Ia peak luminosity. They argued that a change in $G$ would primarily alter the mass of the ejecta $M_{\textrm{ej}}$ while also assuming that the variability in total $M_{\textrm{Ni}}$ is unchanged\footnote{This assumption is probably the most questionable one of the paper as the authors themselves admit. Their semi-analytic model can not predict how much $M_{\textrm{Ni}}$ is produced in type Ia~SNe. If the typical $M_{\textrm{Ni}}$ is altered by a change in $G$ this would affect their final scaling relation between standardized SN luminosity and $G$.}. Assuming $M_\textrm{ej}\propto M_c \propto G^{-3/2}$ implies that an increase of $G$ would lead to a decrease in $M_\textrm{ej}$. A decreased ejecta mass would create a lower density medium around the SNe Ia, increasing the peak luminosity and decreasing the width of the light curve. Upon standardizing the light curves by applying a stretch factor to match the shape of the standard template, they find that they need to increase the width and therefore increase the peak luminosity of their stretched light curves further. They found that light curve standardization would still hold to a good approximation but the peak luminosity of the standardized light curve would be larger when $G>G_N$.

Ref.~\cite{Sakstein2019} performed a fit to the results of \cite{Wright2018} and found a scaling relation for the type Ia SNe true standardized luminosity with $G$ of the form  $L \propto G^{1.46}$, which would correspond to
\begin{equation}
\label{eq:SNeIa_L_G}
L\propto M_c^{-0.97},
\end{equation}
i.e. the standardized luminosity \textit{decreases} with the Chandrashekhar mass.

Given the various possibilities that we have discussed for the scaling of $L$ with $M_c$, we adopt a flexible ansatz in this work and assume that $L \propto M_c^n \propto G^{-3n/2}$.

If the hypothesis of a $G$-transition at some distance $d_T$ is correct, this would imply that there are \textit{two different} standardized SNe peak luminosities. We denote as $L_1$ the standardized peak luminosity for the set of SNe with $d<d_T$, and we denote as $L_2$ the same for SNe with $d>d_T$.

The difference between these two standardizations is then given by,
\begin{equation}
\label{eq:SNLcorr}
 (\textrm{log } L_2)- (\textrm{log } L_1) = -\frac{3 n}{2} \textrm{log} \left (1 + \frac{\Delta G}{G_{\textrm{N}}} \right).
\end{equation}

Since the standardization of SNe~Ia light curves is performed with the Hubble flow SNe at distances greater than 80~Mpc, and no evolution in the properties of the standardized light curves is seen in the Hubble flow SNe, the $G$-transition must occur at distances $d_T$ which are less than $80$~Mpc. At such distances, given the low number of SNe with well calibrated distance measures, it may be that the possibility that SNe light curves fall into distinct classes to the left and to the right of the transition may have escaped detection, especially for sufficiently small values of $\frac{\Delta G}{G}$.

For $n<0$, as in \cite{Wright2018}, we would find that $L_2>L_1$ for $\Delta G >0$. If we incorrectly assumed that the same standardized peak luminosity was valid at all distances, we would therefore underestimate the SNe peak luminosity in the Hubble flow in this situation.

\subsubsection{Combined effect of Cepehids and type Ia SNe on the inferred value of the type Ia SNe standardized peak luminosity}
\label{sec:combined_effect_M}

The key assumptions of the calibration of the peak magnitude of type Ia SNe are i) that the Cepheid PLR is valid at all distances and ii) that there is only one true value of the standardized type Ia SNe peak luminosity which is also valid at all distances. Both these assumptions are violated if there is a $G$-transition at a distance $d_T$ between 7 - 80 Mpc, which lies in the set of calibrator galaxies.

Recall our assumptions that $\Delta G$ is positive. We will additionally assume that the peak SN luminosity $L$ scaling with $M_c$ has index $n<0$. Let us consider two cases for calibrator galaxies below assuming that such a transition in $G$ has taken place:
\begin{itemize}
\item We would find that a Cepheid calibrator to the right of the transition (in a stronger $G$ environment) would have a distance which is underestimated~(eq.~\ref{eq:cepheidcorr}) if one does not take into account the modified intercept of the PLR at these larger distances. This inference would in turn lead to the SN standardized peak luminosity being underestimated.
\item In case the Cepheid calibrator lies to the left of the transition (in a standard $G$ environment). The distances to such Cepheids would be correctly inferred. We could then use this to infer the luminosity of SNe in the calibrator galaxy. However, when applying this inferred luminosity to Hubble flow SNe which lie to the right of the transition (in the stronger $G$ environment), if we assume that the same SNe standardized peak luminosity is valid, we would once again underestimate the luminosity (eq.~\ref{eq:SNLcorr}).
\end{itemize}

Given that we choose our transition distance to lie in the calibrator rung, either of the two cases above might hold for a given calibrator galaxy. Since both cases change the inferred value of the standardized peak luminosity in the same direction, we see that the net effect is that we would underestimate the luminosity, or equivalently overestimate the standardized type Ia SNe peak absolute magnitude (which we will denote as $M$).

\subsubsection{Effect of a $G$-transition on the Hubble flow SNe luminosity distance-redshift relation and inference of the Hubble constant}
\label{sec:SNedlzrelation}

Finally, we discuss the effect of the $G$-transition on the distance-redshift relation. This effect is potentially of importance when applied to Hubble flow SNe to finally determine the Hubble constant once the type Ia SNe peak absolute magnitude $M$ has been calibrated.

The luminosity distance to Hubble flow SNe is given by $d_L(z) = (1+z) r(z)$. Since the SNe in the Hubble flow all lie at redshifts $z>0.02$, which are beyond the transition redshift $z_T$ by assumption, we can use the comoving distance-redshift relation $r(z)$ given in eq.~\ref{eq:rzhighGtrans} for the $G$-transition cosmology,
\begin{eqnarray}
r(z) = \frac{c}{H_{0\beta}} \left [ I(z) + k \right ],
\end{eqnarray}
where the constant $k$ is given in eq.~\ref{eq:kdef}, and the integral $I(z)$ is given by,
\begin{eqnarray}
 I(z) &=& \int_0^z dz^\prime f(z^\prime).
\end{eqnarray}
To proceed further, we can make a Taylor series expansion of $f(z)$ (eq.~\ref{eq:fz}) in the small $z$ limit. We can ignore the radiation component at late times and thus we have,
\begin{eqnarray}
f(z) \approx \frac{1}{\left(\Omega_m (1+z^\prime)^3+   \Omega_\Lambda \right)^{1/2}} &=& 1 -\frac{3 \Omega_m}{2}z + \left[-\frac{3 \Omega_m}{2} + \frac{27 \Omega^2_m}{8} \right] z^2 + \mathcal{O}(z^3), \nonumber \\
&=& 1 + a_0z +a_1z^2 +a_2 z^3 + ...,
\end{eqnarray}
 where $a_0 = -\frac{3\Omega_m}{2}, a_1 =-\frac{3 \Omega_m}{2} + \frac{27 \Omega^2_m}{8}, a_2 = \frac{1}{16} \left[ -8 \Omega_m + 108 \Omega_m^2 - 135 \Omega_m^3 \right] ...$

Thus, $I(z)$ can also be approximated as,
\begin{eqnarray}
 I(z) &=& \int_0^z \frac{dz^\prime}{\left(\Omega_m (1+z^\prime)^3+   \Omega_\Lambda \right)^{1/2}},\\
 &=& z + \frac{a_0}{2} z^2 + \frac{a_1}{3} z^3 +  \frac{a_2}{4} z^4 + ... .
\end{eqnarray}
Note that $k$ can also be expressed in terms of the integral $I(z)$ evaluated at $z_T$ as,
\begin{equation}
k =  \left \{ \left ( 1 + \frac{\Delta G}{G_N} \right  )^{1/2} - 1 \right \} I(z_T).
\end{equation}

We can finally substitute the series representation of $I(z)$ in to the expression for $d_L(z)$ to obtain,
\begin{eqnarray}
\label{eq:dLzGtrans1}
d_L(z) &=& \frac{c}{H_{0\beta}} \left [ k(1+z) +  z + \left( \frac{2+a_0}{2} \right )z^2 + \left( \frac{3 a_0+2a_1}{6} \right )z^3 + ... \right ].
\end{eqnarray}
This expression is valid for $z>z_T$, and by assumption since $z_T$ lies in the calibrator rung this is valid for all Hubble flow supernovae.

This should be compared to the expression for $d_L(z)$ in the standard cosmology which can be obtained by setting $H_{0\beta} \rightarrow H_0$ and $k\rightarrow0$ which gives,
\begin{eqnarray}
d_L(z) &=& \frac{c}{H_0} \left[ z + \left( \frac{2+a_0}{2} \right )z^2 + \left( \frac{3a_0+2a_1}{6} \right )z^3 + ... \right ],
\end{eqnarray}
which can further be matched to the effective low redshift expansion used by the SH0ES collaboration~\cite{R22},
\begin{eqnarray}
\label{eq:dLzLCDM}
d_L(z) &=& \frac{c}{H_0} \left[ z + \left( \frac{1-q_0}{2} \right )z^2 - \left( \frac{1-q_0-3q_0^2+j_0}{6} \right )z^3 + ... \right ].
\end{eqnarray}
Comparing these two expressions we see that,
\begin{eqnarray}
q_0 &=& - 1 - a_0 = -1 + \frac{3\Omega_m}{2},\\
j_0 &=& 1 + 2a_0 +3 a_0^2 - 2a_1 =1,
\end{eqnarray}
which are the standard expectations of $\Lambda$CDM cosmology.

Thus, in the cosmology with a $G$-transition, we can rewrite the expression of eq.~\ref{eq:dLzGtrans1} using the parameters $q_0$ and $j_0$ as,
\begin{eqnarray}
\label{eq:dLzGtrans2}
d_L(z) &=& \frac{c}{H_{0\beta}} \left [ k(1+z) +  z + \left( \frac{1-q_0}{2} \right )z^2 - \left( \frac{1-q_0-3q_0^2+j_0}{6} \right )z^3 +  ... \right ].
\end{eqnarray}
Comparing the above expression with eq.~\ref{eq:dLzLCDM}, we see that the expression for luminosity distance is modified by the same two effects that we had identified in sec.~\ref{sec:CMBinference}, namely the replacement of $H_0$ by $H_{0\beta}$, and the $k$ dependent corrections.

In particular the new form of the $d_L(z)$ relation implies an important result which is that the Hubble law of eq.~\ref{eq:Hubblelawstd} is modified in the presence of a $G$-transition at low $z$ (but still $z>z_T$) to be approximately,
\begin{eqnarray}
\label{eq:dLzGtransapprox}
d_L(z) &\simeq& \frac{cz}{H_{0\beta}} \left [ 1+ \frac{k}{z}+ k   \right],
\end{eqnarray}
where the terms proportional to $k$ parameterize the deviation of the Hubble law (with constant $H_{0\beta}$) from the standard prediction\footnote{Since we have an FRW metric even in the case of the $G$-transition cosmology, it seems surprising that the Hubble law doesn't work. Actually, the Hubble law still holds for $z<z_T$ with $d_L(z) \simeq \frac{cz}{H_{0\alpha}}$, however at such low redshifts there are no Hubble flow SNe through which one could test this relationship and also measure the constant $H_{0\alpha}$.}. We will argue that these $k$ dependent corrections are small for the SNe used to infer the Hubble constant in the SH0ES analysis and thus, the effective Hubble constant that one can extract from the data would correspond simply to $H_{0\beta}$.

As we have mentioned earlier, in order to correctly infer $H_{0\beta}$ from the data, one needs to correctly take into account the corrections to the SNe and Cepheid behaviour beyond the transition distance $z_T$. Since the SH0ES team assumed the standard cosmology and not the $G$-transition cosmology, they assumed that the SNe and Cepheid behavior is the same at all distances.

Let us review the analysis of the SH0ES~'22~\cite{R22} which uses the expression for $d_L(z)$ from the standard cosmology as given in eq.~\ref{eq:dLzLCDM}. For a supernova of calibrated absolute peak magnitude $M$ at a redshift $z$, the apparent magnitude is given by,
\begin{equation}
\label{eq:SNmdL}
m = M + 5 \textrm{ Log } \frac{d_L(z)}{10~\textrm{pc}}.
\end{equation}
Thus we get,
\begin{equation}
\label{eq:SNezm}
m = M + 25 - 5 \textrm{ Log }\frac{H_0}{\textrm{km/s/Mpc}}   + 5 \textrm{ Log } \frac{c z}{\textrm{km/s}} \left \{ 1  + \frac{1}{2}(1-q_0)z - \frac{1}{6} \left(1-q_0 - 3q_0^2 + j_0  \right ) z^2 + \mathcal{O}(z^3) \right \}.
\end{equation}
 This is the relationship used by~\cite{R22} to determine the Hubble constant. First, note that the non-trivial redshift dependence means that this relationship can be applied to far away SNe, deeper in the Hubble flow where the corrections to the simple Hubble law of eq.~\ref{eq:Hubblelawstd} are important. Fitting the form of the observed SNe apparent-magnitude redshift relation be used to yield the values of $q_0$ and $j_0$. The SH0ES team find that $q_0= -0.55$ and $j_0=1$, respectively. Note that the value of $q_0$ implies a value of $\Omega_m = 0.3$, which is in agreement with the value inferred from the CMB. The \Rsix analysis~\cite{R16} claims that the uncertainty on $q_0$ contributes only an $\mathcal{O}(0.1\%)$ uncertainty in the determination of the Hubble constant, which is a minor effect compared to the percent level uncertainty on the value of $H_0$ that they obtain.

Then this relationship is applied to 277 lower redshift SNe between $0.023 < z < 0.15$. Combined with knowledge of the calibrated absolute magnitude $M$ of the SNe from calibrators, \Rtwo~\cite{R22} determine the intercept of the $m(z)$ relation for these low redshift SNe. This yields the intercept $a_B$ which is defined as,
\begin{equation}
a_B =  \textrm{ Log } \frac{c z}{\textrm{km/s}} \left \{ 1  + \frac{1}{2}(1-q_0)z - \frac{1}{6} \left(1-q_0 - 3q_0^2 + j_0  \right ) z^2 + \mathcal{O}(z^3) \right \}  - 0.2 m.
\end{equation}
At low redshifts this is approximately given by, $a_B =  \textrm{ Log } \frac{c z}{\textrm{km/s}} -0.2 m$.

The SH0ES~'22~\cite{R22} analysis finds a value of $a_B =0.714158$. The uncertainty on the extracted value of $a_B$ is $\mathcal{O}(0.1\%)$~\cite{R16} and is once again a subdominant source of uncertainty on the value of the Hubble constant that they are able to finally extract.

Using this value of $a_B$, and the calibrated SNe peak magnitude, the Hubble constant assuming the standard cosmology can be determined as,
\begin{equation}
\label{eq:H0riess}
\textrm{ Log }\frac{H_0}{\textrm{km/s/Mpc}} = \frac{M + 25 + 5 a_B}{5}.
\end{equation}
The dominant contribution to the percent level uncertainty on the extracted Hubble constant comes from the uncertainty on the calibrated absolute peak magnitude $M$, which is determined from the distance ladder. The $0.1\%$ error on the parameter $a_B$ leads to a subdominant (0.1\% level) source of uncertainty on $H_0$.

The relationship of eq.~\ref{eq:H0riess} also makes clear a neat separation in the determination of the Hubble constant into obtaining two intermediate constants which depend on distinct data sets i)~$M$~from calibrator SNe and Cepheids, and ii)~$a_B$~which is determined from Hubble flow SNe. The $d_L(z)$ relationship only affects the determination of $a_B$ from Hubble flow SNe, but it does not affect the determination of $M$ since we do not use redshift information for SNe in the calibrator galaxies.

\vspace{5mm}
How is this inference altered in a cosmology with a $G$-transition? First we can plug in our modified expression for the luminosity distance eq.~\ref{eq:dLzGtrans2} into the apparent magnitude relation for SNe eq.~\ref{eq:SNmdL}.
This gives,
\begin{equation}
\label{eq:SNezmGtrans}
m = \widetilde{M} + 25 - 5 \textrm{ Log }\frac{H_{0\beta}}{\textrm{km/s/Mpc}}   + 5 \textrm{ Log } \frac{c z}{\textrm{km/s}} \left \{ 1 + \frac{k}{z} + k  + \frac{1}{2}(1-q_0)z - \frac{1}{6} \left(1-q_0 - 3q_0^2 + j_0  \right ) z^2 + \mathcal{O}(z^3) \right \}.
\end{equation}
Here, we use a different symbol for the calibrated SNe peak magnitude $\widetilde{M}$ than that of the standard analysis $M$, since as discussed in the previous subsection the SNe peak magnitude will in general be different for distant SNe and thus needs to be recalibrated.
The relationship between apparent magnitude and redshift given in eq.~\ref{eq:SNezmGtrans} above is similar to the law in eq.~\ref{eq:SNezm} for the standard cosmology, with the following changes --- (i) the replacement of $H_0$ by $H_{0\beta}$ (ii) $k$ dependent corrections, and (iii) the replacement of $M$ by $\widetilde{M}$.

Now for the distant supernovae, which are at redshift $z \gtrsim 1$ and are used to determine $j_0$ and $q_0$, the $k$ dependent corrections are suppressed by a factor of $O\left(\frac{\Delta G}{G} \frac{z_T}{z} \right) \lesssim 0.1 \times \frac{0.01}{1} \simeq 10^{-3}$ based on our assumptions on the parameter range of the $G$-transition cosmology laid out in sec.~\ref{sec:assumptions}.
Thus, even if the cosmology is that of the $G$-transition, and one were to fit for $q_0$ and $j_0$ assuming the incorrect relationship of eq.~\ref{eq:SNezm}, we expect that it would only have a 0.1\% shift in the extraction of these parameters. The shift in the value of $q_0$ is of the order of the uncertainty on $q_0$ from the fit, which as we have already mentioned is a subdominant source of uncertainty on the finally extracted Hubble constant. Thus, we can simply assume that the values of $q_0$ and $j_0$ that would be extracted by redoing a fit assuming the appropriate relationship of eq.~\ref{eq:SNezmGtrans} would be identical to those of the standard cosmology, i.e. $q_0= -0.55$ and $j_0=1$.

Now one can use the SNe between $0.023 < z < 0.15$ to obtain a modified $a_B$ which we denote as $\widetilde{a}_B$,
\begin{equation}
\widetilde{a}_B =  \textrm{ Log } \frac{c z}{\textrm{km/s}} \left \{ 1 + \frac{k}{z} + k + \frac{1}{2}(1-q_0)z - \frac{1}{6} \left(1-q_0 - 3q_0^2 + j_0  \right ) z^2 + \mathcal{O}(z^3) \right \}  - 0.2 m,
\end{equation}
which at low redshift (but still $z>z_T$) can be approximated as $\widetilde{a}_B \simeq \textrm{ Log } \frac{c z}{\textrm{km/s}}\left ( 1+ \frac{k}{z} + k \right) -0.2 m$. Once again the modifications to the equation for $\widetilde{a}_B$ are suppressed by a factor $O\left(\frac{\Delta G}{G} \frac{z_T}{z} \right)$. However, now the redshift $z$ of the SNe under consideration are smaller. The SH0ES team has not released a full table of the redshift of these SNe, although they have described how the selection of this data can be made from the Pan-STARRS catalog~\cite{Pan-STARRS1:2017jku}. If we assume a typical redshift $z \simeq 0.1$ of SNe in this sample, the $k$ dependent terms are suppressed by a factor of $O\left(\frac{\Delta G}{G} \frac{z_T}{z} \right) \lesssim 0.1 \times \frac{0.01}{0.1} \simeq 10^{-2}$, i.e. this could lead to an $\mathcal{O}(1\%)$ shift in the value of $\widetilde{a}_B$ away from $a_B$. The size of this shift would be smaller if $\Delta G$ or $z_T$ are not near the upper end of the parameter range that we are considering.

We thus expect that a reanalysis of the Hubble flow SNe between $0.023 < z < 0.15$ would yield a modified value of $\widetilde{a}_B$ which would be offset from the value of $a_B =0.714158$ found by ~\cite{R22} assuming the standard cosmology, by a correction of $O\left(\frac{\Delta G}{G} \frac{z_T}{z} \right)\lesssim 0.01$. Additionally, if the corrections are sufficiently large, then the deviation of the form of the Hubble law from its standard form (specifically because of the $k$ dependent corrections) may also be testable with the low redshift SNe sample.

Once we know both the parameters $\widetilde{a}_B$ from the Hubble flow SNe, and $\widetilde{M}$ from the distance ladder calibration of SNe peak luminosities, we can infer the Hubble constant $H_{0\beta}$ of the $G$-transition cosmology as,
\begin{equation}
\label{eq:H0Gtransition}
\textrm{ Log }\frac{H_{0\beta}}{\textrm{km/s/Mpc}} = \frac{\widetilde{M} + 25 + 5 \widetilde{a}_B}{5}.
\end{equation}

\vspace{3mm}
\textbf{Effect of a $G$-transition on the inference of the Hubble constant}

Assuming that the $G$-transition cosmology is correct (with $\Delta G/G>0$) and given the correct relationship for the Hubble constant is given by eq.~\ref{eq:H0Gtransition}, we can ask what error we would make on the inferred value of the Hubble constant if, proceeding as the SH0ES team did, we assume instead eq.~\ref{eq:H0riess} which is valid in the standard cosmology?

We see that there are two sources of error. The first error, as noted in sec.~\ref{sec:combined_effect_M}, is that we would have incorrectly overestimated the type Ia SNe peak absolute magnitude, i.e. $M > \widetilde{M}$ and this effect would be $\mathcal{O}\left(\frac{\Delta G}{G} \right)$. However, the second error, is that we would also have underestimated the value of $a_B < \widetilde{a}_B$ with the difference being $\mathcal{O}\left( \frac{\Delta G}{G} \frac{z_T}{z_\textrm{SN}} \right)$, where $z_\textrm{SN} \simeq 0.1$ is the typical redshift of the low redshift SNe sample in the Hubble flow. Since the first effect is more dominant, we see that this would lead to an overestimate of the inferred value of the Hubble constant and this could potentially explain the discrepancy with the Hubble constant extracted from CMB observations.

\vspace{3mm}
Assuming that a $G$-transition took place, if we want to correctly infer the Hubble constant (in this case $H_{0\beta}$) of the $G$-transition cosmology, we would need to reanalyze the data and (i) recalibrate the type Ia SNe peak magnitude to obtain $\widetilde{M}$, and (ii) refit the Hubble flow SNe data to obtain the value of $\widetilde{a}_B$. We will outline in the next section how this can be done.

\section{Methodology - fitting low redshift data to a $G$-transition hypothesis}
\label{sec:analysis}

In section \ref{sec:intuition}, we presented a theoretical overview of how the standard distance ladder is built using Cepheids as calibrators for type Ia SNe, and how once this calibration is accomplished, Hubble flow SNe can be used to extract the Hubble constant. We also explained how if there was a $G$-transition at some time corresponding to a look-back distance $d_T$ between 7 - 80~Mpc, this would lead to an incorrect inference of the Hubble constant using local universe observations.

Our goal is to see whether such a $G$-transition is actually \textit{preferred} by the observed data over the standard hypothesis of no $G$-transition. There are two ways in which we can potentially look for evidence of a $G$-transition. The first is to look at SNe in calibrator galaxies and look for evidence for a transition in the standardized luminosity of SNe in the calibrators. The second is to look for evidence of a modified Hubble law, eq.~\ref{eq:dLzGtransapprox} which would be most significant for low redshift SNe in the Hubble flow. While both effects have an $\mathcal{O}(\frac{\Delta G}{G})$ suppression, the latter effect has an additional suppression by a factor of $\frac{z_T}{z_{SN}}$. Hence, we will direct our focus to the first method.

As we shall see, we cannot obtain the value of $\Delta G$ from the Cepheid and SNe data, however we can still look for evidence of a transition in the standardized luminosity of SNe in calibrator galaxies and we can find the best-fit transition distance. If we find that this preferred transition distance is such that $\frac{z_T}{z_{SN}}$ is sufficiently small, then the modifications to the form of the Hubble law are expected to be practically unobservable in Hubble flow SNe, and the value of $\widetilde{a}_B$ and $a_B$ will be nearly identical. Furthermore, given this preferred transition distance, we can then find a value of $\Delta G$ such that the standardized SN luminosity for calibrators to the right of the transition (and also that of Hubble flow SNe) is such that it leads to a Hubble constant which is in agreement with the value obtained from CMB data.

To accomplish our goal, we need to fit the empirical data from Cepheids and SNe to both of these hypotheses and compare the quality of the fits. This data is described in sec.~\ref{sec:data}.

We outline below our analysis procedure.
\begin{itemize}
\item For the standard hypothesis of no $G$-transition, we will attempt to reproduce analysis of SH0ES'22~\cite{R22}. However our analysis will make several simplifying assumptions which differ from their more detailed analysis. In order to establish the validity of our procedure, we will show that we obtain a value of the Hubble constant which is in good agreement with that of~\cite{R22}, which establishes confidence in our simplified procedure. This procedure and the results will be discussed in sec.~\ref{sec:standardfit}.
\item We will then go on to describe how we modify this analysis to include the hypothesis of a $G$-transition. In order to specify the alternative hypothesis - we need to specify the $G$-transition parameters $\Delta G$ and $d_T$. We leave the supernova standardized luminosity $L-M_c$ scaling index $n$ as a derived fit parameter by allowing for a different standardized SN peak luminosity to the left and to the right of the transition. We do not impose any prior on the sign of $n$. The procedure and results of this step will be discussed in sec.~\ref{sec:Gtransitionfit}.
\end{itemize}

For each hypothesis above, we can compute the derived value of the Hubble constant ($H_0$ or $H_{0\beta}$) using the appropriate relation, eq.~\ref{eq:H0riess} in the standard scenario, or the analogous eq.~\ref{eq:H0Gtransition} for the $G$-transition scenario.

The calibrator SNe data can be used to find the standardized SN peak absolute magnitude of SNe in the Hubble flow. The inference of this value will differ depending upon whether we assume a hypothesis of a $G$-transition or no $G$-transition (in particular for the hypothesis of the $G$-transition, we must use the standardized absolute magnitude to the right of the transition).

In principle, to determine the Hubble constant in the $G$-transition hypothesis, we also need to refit the low redshift Hubble flow SNe data to determine the value of $\widetilde{a}_B$, which is the intercept of the $B$-band apparent magnitude - redshift relation. However, if we find that the best-fit value of $d_T$ is sufficiently small, then we can simply take $\widetilde{a}_B=a_B$ (we will indeed find this to be the case).

For the $G$-transition hypothesis, we can also compute the inferred value of the SN scaling index~$n$.

After fitting both sets of hypotheses, we can then compare the quality of the fits --  while appropriately penalizing for the extra parameters in the $G$-transition hypothesis -- to compare which model provides a better fit to the data. We present this comparison in sec.~\ref{sec:comparison}.

\section{Description of the Cepheid and SNe observational data set used}
\label{sec:data}
We will use data from the \Rtwo \cite{R22} analysis. For our purposes, we will only need to fit the SNe data in the calibrators to either the hypothesis of a $G$-transition or no $G$-transition, thus it will be sufficient to use data from Table 6 of \cite{R22}. This table presents fitted distance moduli to 37 calibrator galaxies from Cepheids in anchors and calibrator galaxies along with their uncertainties\footnote{The table actually presents two different sets of distances to calibrator galaxies, those inferred from a simultaneous fit including SNe data and another set which does not include SNe data. For our purposes, we will be interested in the latter set which are extracted independently of the SNe data.}. In addition, the table also contains the observed $B$-band peak apparent magnitudes for 42 type Ia SNe that lie in these calibrator galaxies\footnote{Three galaxies have two separate SNe each and another galaxy has three separate SNe.}.

For Hubble flow SNe, we use the value of $a_B$= 0.714158 as determined in \Rtwo~\cite{R22}. This value was determined by using 277 Hubble flow SNe Ia at redshifts $0.023<z<0.15$, assuming the standard distance-redshift relation.

\section{Fitting the distance ladder assuming no $G$-transition}
\label{sec:standardfit}
We first present the methodology used to fit the distance ladder without a $G$-transition, i.e. in the standard scenario. Our procedure is similar in spirit to that of \Rtwo \cite{R22}, with a few simplifying assumptions.

We make use of the fitted distances to the calibrator galaxies from \cite{R22} which are obtained from a combination of Cepheid data in the anchor and calibrator galaxies without repeating this part of the analysis. Using these distances in conjunction with the apparent magnitudes of SNe observed in these hosts, we then obtained a fitted value for the  standardized absolute magnitude of SNe~Ia (denoted as $M_B$).

This value of $M_B$ can then be used along  with the observed apparent magnitudes of Hubble flow SNe to obtain the distances to their host galaxies, and this can be further used to infer the value of the Hubble constant.

For the SNe Ia in the calibrator galaxies we have,
\begin{equation}
\label{eq:SNeIaMB}
    m_{B,i} = \mu_{i} + M_B,
\end{equation}
where $m_{B,i}$ is the observed $B$-band peak apparent magnitude after application of the light curve shape fitting correction for a type Ia SN in a calibrator galaxy and $\mu_i$ is the already fitted distance (from Cepheids) to the galaxy. Here the parameter $M_B$, which is the standardized $B$-band absolute magnitude, is to be extracted from a fit to the data.

In what follows, we describe how we obtain the best fit value of the peak absolute magnitude $M_B$ in the calibrator galaxies. Here, the value of $M_B$ is assumed to be the same for all SNe Ia. After fitting the data to obtain $M_B$, we can use this value in eq.~\ref{eq:H0riess} to infer the value of the Hubble constant.

\subsection{Fit using $\chi^2$ minimisation}
\label{sec:chi2_standard}
We fit the observed SNe apparent magnitudes $m_{B,i}$ to obtain the SNe Ia standardized peak absolute magnitude $M_B$ in calibrator galaxies (eq.~\ref{eq:SNeIaMB}). To do this, we first define a $\chi^{2}$ or equivalently a log-likelihood (where $\chi^{2}= -2 \log\mathcal{L}$) and perform a minimization over all possible values of $M_B$.

Our total $\chi^2$ is simply defined as,
\begin{equation}
\label{eq:LSN}
   \chi^2 = \sum_{i,j}  \left ((m_{B,i})^{\rm{obs}} - (m_{B,i})^{\rm{model}} \right) \, {[C^{-1}]}_{ij} \, \left((m_{B,j})^{\rm{obs}} - (m_{B,j})^{\rm{model}} \right),
\end{equation}
where $m_{B,i}$ is the corrected $B$-band peak apparent magnitude of the $i$-th SNe Ia. The superscript \enquote{obs} corresponds to the observed value of $m_{B,i}$ and the superscript \enquote{model} corresponds to the value of $m_{B,i}$ calculated from the theoretical model of the standard distance ladder (equation \ref{eq:SNeIaMB}). In the equation
${[C^{-1}]}_{ij}$ are elements of the inverse covariance matrix. The covariance matrix $C$ has diagonal entries which are given by $C_{ii} = \delta m_{B,i}^2 + \delta \mu_i^2 $, i.e. the quadrature sum of the apparent magnitude and distance modulus errors for a SNe in a given calibrator. However, the covariance matrix also has non-zero off-diagonal entries given by, $C_{ij} = \delta \mu_i^2 $, if the $i$-th and $j$-th SNe lie in the same galaxy.

Extremizing the final chi-squared function, we obtain the best fit value and 1-$\sigma$ confidence intervals on the parameter $M_B$\footnote{The 1-$\sigma$ interval on $M_B$ is obtained by the inversion of the Hessian as, $\sigma_{M_B} = \sqrt{\frac{2}{\frac{\partial^2 \chi^2}{\partial M_B^2}}}$.}.

\subsection{Results and validation of fit to the distance ladder}
\label{sec:resultsnoG}

After performing the fit described above, we obtained a minimum $\chi^2$ of 40.1 for 42 data points with one free parameter $(M_B)$. This gives us a $\chi^2$ per degree of freedom ($\chi^2_{\textrm{dof}}$) of $0.98$.

Our best fit value of $M_B = -19.26~\pm~0.02$. By substituting the obtained value of $M_B$ in the $H_0-M_B$ relation (eq.~\ref{eq:H0riess}), we obtain an inferred value of the Hubble constant $H_0~=~72.83~\pm~0.71$~\Hunit. The value of the Hubble constant that we infer is consistent with that of \Rtwo~\cite{R22} of $H_0~=~73.04~\pm~1.04$~\Hunit. This value is based on a slightly different analysis strategy than ours, where the authors perform a simultaneous fit to all the Cepheid and SNe data and they also correctly include a systematic error on $H_0$. Nonetheless, our inferred value is in good agreement with the values obtained by \Rtwo and validates our simplified analysis procedure.

In the next section, we begin by discussing how to modify this analysis by taking into account a $G$-transition, and then we perform a fit to the data for this alternate hypothesis.

\section{Fitting the distance ladder to a $G$-transition}

\label{sec:Gtransitionfit}

While performing a fit to the distance ladder \Rtwo \cite{R22} obtained distance moduli to the 37 Cepheid galaxies that range from $\mu = 29.2$ to $\mu =34.5$ which correspond to luminosity distances between $7$ and $80$~Mpc.

We now discuss the hypothesis of a $G$-transition at a lookback time corresponding to a distance modulus $\mu_T$ (or a distance $d_T$) which lies in this range. We will assume that $G$ was larger than $G_N$ in the past, by an amount $\Delta G$. Cepheids that are at distances smaller that $d_T$ will have distances which are correctly inferred, but Cepheids that lie at distances larger than $d_T$ will have underestimated distances.

Given a hypothesis with specific values of $\mu_T$ and $\Delta G$, we would therefore find that the corrected distances to the Cepheid calibrator galaxies, which we denote as $\mu^\prime_i$, are given by
\begin{equation}
\mu^\prime_i = \begin{cases}
\mu_i,  \textrm{if $\mu_i < \mu_T$}\\
\mu_i + \mu_{\textrm{corr}}, \textrm{if $\mu_i > \mu_T$},  \\
\end{cases}
\end{equation}
where the distances $\mu_i$ are the distance moduli found by \Rtwo and the distance correction factor $\mu_{\textrm{corr}}$ is given by (see eq.~\ref{eq:cepheidcorr} and eq.~\ref{eq:PLR_intercept_G_1}),
\begin{equation}
\label{eq:cepheidmucorr}
\mu_{\textrm{corr}}= 2.5 \left(\frac{\alpha}{2} + B \right) \textrm{Log} \left( 1 + \frac{\Delta G}{G} \right),
\end{equation}
where $\alpha$ is the PLR slope and the coefficient $B$ depends on the Cepheid mass and crossing of the instability strip (see sec.~\ref{sec:cep_PLR_G}). The value of $\alpha$ is obtained from the fit of \Rtwo as $\alpha = 1.32$\footnote{Recall from our discussion in sec.~\ref{sec:cep_PLR_G}, that the slope of the PLR is unaffected by the $G$-transition, and hence this value should remain unchanged from the fit to the no $G$-transition case performed in \Rtwo. \Rtwo actually report a fitted value of $b_W = -3.299\pm0.015$~\cite{R22}, where $b_W=-2.5\alpha$ is the slope of the PLR when using absolute magnitude rather than luminosity. We also ignore the small uncertainty on $b_W$ which is subdominant compared to the uncertainty on~$B$.}. In our analysis, we take a fiducial value $B = 4$. Though later in this section, we also discuss how our results may change when we take the extreme values of $B$ found in Sakstein et al. \cite{Sakstein2019}. For positive $\Delta G$, the distance modulus correction $\mu_{\textrm{corr}}$ is also positive, thus Cepheids beyond the transition distance $\mu_T$ lie further away than the distances inferred by \Rtwo.

Note that the observable for Cepheids is the apparent magnitude $m$, which is theoretically calculated as $m = M + \mu_i$. The best-fit prediction for the apparent magnitude is unchanged even in the presence of a $G$-transition since the change in the distance modulus compensates the change in the intrinsic Cepheid absolute magnitude. Thus, the quality of fit parameter which is the minimum $\chi^2$ of the \Rtwo fit to the Cepheid variables is completely unaffected even when correctly accounting for a $G$-transition.

However, once we correct the Cepheid inferred distance moduli using eq.~\ref{eq:cepheidmucorr}, we would then use these corrected distances to predict the SNe apparent magnitudes,
 \begin{equation}
 \label{eq:SNe_with_G}
    m^{\textrm{pred}}_{B,i} = \begin{cases}
\mu^\prime_i + \widetilde{M}_{B1},  \textrm{if $\mu_i < \mu_T$}\\
\mu^\prime_i + \widetilde{M}_{B2}, \textrm{if $\mu_i > \mu_T$},  \\
\end{cases}
    \end{equation}
where we have introduced two different absolute magnitude parameters $\widetilde{M}_{B1}$ and $\widetilde{M}_{B2}$ to denote the standardized type Ia SNe peak brightness to the left and to the right of the transition, respectively.
Note that the correction to the distance moduli $\mu_{\textrm{corr}}$ directly adds to the parameter $\widetilde{M}_{B2}$, thus only the combination $\widetilde{M}^\prime_{B2} = \widetilde{M}_{B2}+\mu_{\textrm{corr}}$ can directly be constrained by observations of type Ia SNe in the calibrators. In other words the value of $\Delta G$ can not be determined from the data set that we are working with. However, if we assume a value of $\Delta G$, this would fix the value of $\widetilde{M}_{B2}$.

The value of the Hubble constant $H_{0\beta}$ can be inferred from our fitted parameters by using eq.~\ref{eq:H0Gtransition} with $\widetilde{M}$ set to $\widetilde{M}_{B2}$, i.e. by using the standardized peak luminosity for distant SNe (to the right of $\mu_T$). We also need to specify the value of $\widetilde{a}_B$ by refitting the Hubble flow SNe data, but if we find that the transition redshift $z_T$ is sufficiently small, then to a good approximation we can set $\widetilde{a}_B$ to $a_B$ taken from the \Rtwo fit.

We do not fix a relationship between $\widetilde{M}_{B1}$ and $\widetilde{M}_{B2}$ when fitting, but rather we allow them to be free fit parameters. This is equivalent to allowing the index $n$ of the SNe~Ia $L-M_c$ relation to be determined from the fit as a derived parameter. Once we obtain the best-fit values of $\widetilde{M}_{B1}$ and $\widetilde{M}_{B2}$, the value of $n$ can be inferred from our fit by inverting eq.~\ref{eq:SNLcorr} as,
\begin{equation}
\label{eq:nvalue}
n = \frac{2}{7.5} \frac{\widetilde{M}_{B2}-\widetilde{M}_{B1}}{\log\left(1 + \frac{\Delta G}{G_N}\right)}.
\end{equation}

\subsection{$\chi^2$ minimisation}
For each hypothesis of $\Delta G$ and $\mu_T$ we can now fit our distance ladder to obtain the fit parameters by minimizing a $\chi^2$. We define our $\chi^2$ exactly the same way as in the case of the no $G$-transition hypothesis (see eq. \ref{eq:LSN}). The only difference now is that for SNe to the left of the transition the apparent magnitude is predicted assuming an intrinsic brightness $\widetilde{M}_{B1}$ and for SNe to the right of the transition, the apparent magnitude is predicted using the effective intrinsic brightness parameter $\widetilde{M}^\prime_{B2}$.

In effect, for a given value of $\mu_T$, this then corresponds to splitting the sum over SNe in the chi-squared into two parts, the first involving SNe in galaxies to the left of the transition, and the second involving galaxies to the right of the transition. Note that this split can always be done since there is no off-diagonal covariance for SNe in different galaxies.
Then the chi-squared for SNe to the left of the transition and chi-squared for SNe to the right of the transition can be separately minimized to find the best fit values of $\widetilde{M}_{B1}$ and $\widetilde{M}^\prime_{B2}$. Adding the chi-squared of each part together gives us the total best-fit chi-squared.

The split can be performed by considering different values of $\mu_T$ that lie between different calibrator galaxies. We consider in turn, all possible splittings on the calibrators, by considering a corresponding discrete set of $\mu_T$ values. For each $\mu_T$ value, we then extremize the $\chi^2$ function to obtain the fitted values of $\widetilde{M}_{B1}$ and $\widetilde{M}^\prime_{B2}$. We also construct the errors on the fitted parameters $\widetilde{M}_{B1}$ and $\widetilde{M}_{B2}$ by using the inverse Hessian. It is easy to see that these parameters have zero correlation since they appear in different terms in the chi-squared sum.

Finally, following this procedure, we choose the $\mu_T$ value that leads to a global minimization of the total chi-squared. This optimal value of $\mu_T$ (up to the nearest calibrator) is the one that is most preferred by the data. However, a precise error on $\mu_T$ can not be given because of the discontinuous dependence of $\chi^2$ on $\mu_T$.

Thus, in effect we are performing a 3 parameter fit to $\widetilde{M}_{B1}$, $\widetilde{M}^\prime_{B2}$ and $\mu_T$. The fit we are performing is essentially to see if two different SNe peak brightnesses provide a better fit to the calibrator data than the assumption of a single peak brightness at all distances.

\subsection{Results for the distance ladder fit in the presence of a $G$-transition}
\label{sec:results}
After optimizing over parameters we find that the minimum $\chi^2$ is 35.7, which yields a chi-squared per degree-of-freedom of 0.91 for our 3 parameter fit.
The best fit parameters are  $\widetilde{M}_{B1} = -19.32 \pm 0.03$, $\widetilde{M}^\prime_{B2} = -19.22 \pm 0.03$, and $\mu_T =31.75 \pm 0.03$. Here, the error we have quoted on $\mu_T$ only indicates the distances to the nearest calibrator galaxy from the transition. The value of $\mu_T$ indicates a best-fit transition at a lookback distance $d_T$ of 22.4 Mpc, or a transition which occurred 73 million years ago. Twelve calibrator galaxies, hosting 14 SNe in the \Rtwo sample lie to the left of this transition.

Let us now further make the choice $\Delta G/G = 0.04$. This choice then fixes $\widetilde{M}_{B2}= - 19.42 \pm 0.03$. The redshift of the transition $z_T$ depends on the inferred value of the Hubble constant, but it can be approximated by inverting the Hubble law to give $z_T \simeq \frac{H_0 d_T}{c} \simeq 0.005$\footnote{Strictly speaking one should use $H_{0\alpha}$ here, but for the level of accuracy we are interested in here it does not matter whether we use $H_0$ from the standard cosmology or $H_{0\alpha}$.}.
In principle now to determine the Hubble constant, one needs to reanalyze the Hubble flow SNe data to obtain $\widetilde{a}_B$. However, given our $z_T$ this implies that $k\simeq \frac{1}{2}\frac{\Delta G}{G_N}\frac{z_T}{z_\textrm{SN}} \simeq \frac{1}{2} \times 0.04 *\frac{0.005}{0.1} \simeq 10^{-3}$ is sufficiently small. Thus, for such a small transition redshift the $k$ dependent corrections would lead to a
difference between $a_B$ and $\widetilde{a}_B$ which is negligible $\mathcal{O}$(0.1\%), and so we can use eq.~\ref{eq:H0Gtransition}, with $\widetilde{a}_B \rightarrow a_B$, to infer the value of the Hubble constant.

This yields $H_{0\beta} = 67.55 \pm 0.83$~\Hunit, which is in good agreement with the Planck inference of $H_0$ (see discussion in sec.~\ref{sec:CMBinference}). Moreover, this choice also fixes the value of the index $n = -1.68 \pm 0.68$ which is also in good agreement with the semi-analytic prediction of ref.~\cite{Wright2018} of $n=-0.97$ at nearly the  $1$-$\sigma$ level.

Note that we have seen in sec.~\ref{sec:CMBinference} that the value of $H_{0\beta}$ inferred from the CMB may be larger than that of $H_0$. Our simple estimate, using the effect of a larger value of $G$ on the sound horizon, indicated that the inferred value of $H_{0\beta}$ would be larger than the Planck value of $H_0$ inferred by assuming the standard cosmology by about 3\% for $\Delta G/G\simeq 4\%$ (see eq.~\ref{eq:deltaH}). Thus it is possible that the Hubble tension may be reconciled with a value of $\Delta G/G$ smaller than the value of $4\%$ that we have assumed here. However, a precise quantification of this value would require a reanalysis of the CMB fits taking into account the presence of a $G$-transition.

\subsection{Effect of the change of $B$ or $m$}
We now comment on the effect on our analysis if we take a different choice of $B$. The first thing to note, is that the value of $\Delta G$ and $B$ appear together in the expression in eq.~\ref{eq:cepheidmucorr}. The second is that our fit is to $\widetilde{M}^\prime_{B2}$, and hence the value of $\Delta G$ is undetermined in our fit. Thus, the effect of a different value of $B$ would show up when we attempt to fix $\Delta G$ to obtain a given value of the Hubble constant -- in our case we demand that we get a value that is consistent with the CMB inference of $H_{0\beta}$. Let us proceed with the assumption stated in sec.~\ref{sec:CMBinference}, that even in the presence of a $G$-transition, the value of $H_{0\beta}$ inferred from the CMB remains at the value of $H_0$ determined by Planck assuming the standard cosmology, i.e. $H_{0\beta} = 67.66 \, \pm  \, 0.42$~\Hunit. Since $\widetilde{M}^\prime_{B2}$ is fixed by our fit, and $\widetilde{M}_{B2}$ is fixed by our demand for the value of the Hubble constant, this can only be achieved by demanding that the value of the distance correction factor ($\mu_\textrm{corr}$ in eq.~\ref{eq:cepheidmucorr}) is unchanged from the value obtained when using $B=4$.

For concreteness, let us consider the two possible extreme values of $B=3.46$ or $B=4.52$ mentioned in Sakstein et al.~\cite{Sakstein2019}. This would change the value of $\Delta G/G$ to 4.5\% or 3.6\%, respectively. This would correspondingly lead to a change in the best fit value of $n$ to $-1.48\pm0.60$ or $-1.86\pm 0.75$, respectively. These values are still broadly consistent with those obtained by Wright and Li~\cite{Wright2018}.

We can also study the effect of relaxing the assumption that the density $\overline{\rho}$ of the Cepheid envelope does not scale with $G$. If we assume that $\overline{\rho}\propto G^m$, where $m\neq 0$ is a scaling index, we would find that the expressions for the distance moduli corrections in eq.~\ref{eq:cepheidmucorr} would be corrected to,
\begin{eqnarray}
\label{eq:cepheidmucorrm}
\mu_{\textrm{corr}} &=& 2.5 \left(\frac{\alpha (1 + m) }{2} + B \right) \textrm{Log} \left( 1 + \frac{\Delta G}{G} \right), \\
&=& 2.5 \left(0.66 + 0.66 m + B   \right) \textrm{Log} \left( 1 + \frac{\Delta G}{G} \right),
\end{eqnarray}
where we have substituted the value of $\alpha=1.32$ in the second line. Thus, we see that the effect of a non-zero scaling index $m$ is equivalent to a change in $B$ as far as the correction to the distance moduli are concerned. Thus, the range of $B$ values that we have considered above would correspond alternatively to a selection of $-0.81 \lesssim m \lesssim 0.79$. For this range of scaling indices, we would find that $\Delta G/G$ between  3.6\% to 4.5\% is needed (for a fixed value $B=4$) to resolve the Hubble tension. If $m$ is larger than $0.79$, smaller values of $\Delta G/G$ will suffice to resolve the Hubble tension. However, if $m \lesssim -0.81$ we will need larger values of $\Delta G/G$.

\section{Model Comparison}
\label{sec:comparison}
We have seen in the previous section that a $G$-transition at $\mu_T =31.75$ provides the best fit to the SNe data. With an additional choice of a 4\% change in $G$, this parameter point also yields a value of $H_{0\beta}$ which is consistent with CMB inferences of the $H_{0\beta}$ parameter.

However, in order to claim that this is a potential solution to the Hubble tension, we must address the key question. Does our preferred $G$-transition model with $\Delta G/G = 4$\% and $\mu_T =31.75$, provide a better fit to the data than the hypothesis of no $G$-transition?

We answer this question by using three estimators of the quality of the fit and comparing their values between these two different hypotheses. The $G$-transition hypothesis has three parameters, viz. the transition distance $\mu_T$, and the two SNe standardized luminosity parameters $\widetilde{M}_{B1}$ and $\widetilde{M}^\prime_{B2}$\footnote{Since there is no way to separate $\Delta G$ and $\widetilde{M}_{B2}$, we take the combination $\widetilde{M}^\prime_{B2}$ as the fit parameter.}, as compared to the single parameter $M_B$ in the no $G$-transition case. Thus, our estimators must penalize for 3 extra parameters of the $G$-transition model.

The first estimator that we use is the $\chi^2$ per degree of freedom, denoted as $\chi^2_{\textrm{dof}}$. The other two are the well-known Akaike Information Criterion (AIC)~\cite{Akaike} and Bayesian Information Criterion (BIC)~\cite{liddle}. To define these estimators, we need the minimum chi-squared $\chi^2_\textrm{min}$, the number of model parameters $d$, and the number of data points $N$. These estimators are then defined as follows,

\begin{eqnarray}
\chi^2_{\textrm{dof}} &=& \frac{1}{N-d} \chi^{2}_{\textrm{min}}, \nonumber \\
\textrm{AIC} &=& \chi^{2}_{\textrm{min}} + 2 d, \nonumber\\
\textrm{BIC} &=& \chi^{2}_{\textrm{min}} + d \ln(N).
\end{eqnarray}

By definition, the model with the lower value of $\chi^2_{\textrm{dof}}$ or AIC or BIC is preferred. We find an improvement in the chi-squared per degree of freedom to 0.91 in the case of a $G$-transition, as opposed to 0.98 in the case without a $G$-transition. We also find an improvement in the AIC, with $\Delta \textrm{AIC} = -0.45$, where the negative sign shows that the AIC has reduced in the case of the hypothesis of a $G$-transition, which indicates a mild preference for this hypothesis. However, in the case of the more stringent BIC criteria which penalizes more strongly for additional parameters, we find that $\Delta \textrm{BIC} = 3.02$, indicating a preference for the standard cosmological model without a $G$-transition.

\section{Summary, discussion and future studies}
\label{sec:discussion}
In this work, we studied the possible effects of $G$-transition between 7 - 80~Mpc on the distance ladder inference of the Hubble constant. We defined the closest analog of the Hubble constant in the $G$-transition cosmology as the parameter $H_{0\beta}$.

We first argued that a reanalysis of CMB data should yield a value of $H_{0\beta}$ in the $G$-transition cosmology which is larger than the value of $H_0$ inferred by assuming the standard cosmology by a factor $0.83\Delta G/G$. However, this result was not computed by a full fit to the CMB data, and is only indicative of the correction that one would obtain after refitting the data.

We then reanalyzed the distance ladder assuming the presence of a $G$-transition, and we found that a $G$-transition at a look-back distance of $\mu_T=31.75$ (22.4 Mpc or 73 Myr) is mildly preferred by the type Ia SNe data when looking at the chi-squared per degree of freedom or AIC criterion, although it is disfavored by the more stringent BIC criterion. If we further assume an effective gravitational constant that was stronger in the past by an amount $\Delta G/G  = 4 \%$, we would then obtain a best fit value of $H_{0\beta} = 67.55 \pm 0.83$~\Hunit from these low redshift probes, which is in excellent agreement with the best-fit value of the Hubble constant $H_0$ as inferred from CMB data~\cite{Planck:2018vyg}. \textit{This would potentially be a resolution of the Hubble tension.} Given that the inference of $H_{0\beta}$ from the CMB might be larger than that of $H_0$, it is possible that a value of $\Delta G/G$ even less than $4\%$ would be sufficient to resolve the Hubble tension. A precise quantification of this value would require a reanalysis of the CMB fits taking into account the presence of a $G$-transition.

In performing our fit to the SNe data, we allowed for the SNe Ia standardized peak luminosity to vary with Chandrashekhar mass as $L\propto M_c^n$, where $n$ is the scaling index. We inferred a best-fit value of $n = -1.68\pm 0.68$, which is in agreement (at nearly the 1-$\sigma$ level) with the theoretical prediction $n=-0.97$ of Wright and Li~\cite{Wright2018}, which used a semi-analytic model for SNe light curves.

The lack of detailed knowledge of the Cepheid parameter $B$, which determines the change in Cepheid luminosity due to a change in $G$, has a relatively minor effect on our analysis, changing the value of the required $\Delta G$ between 3.6\% and 4.5\%.

Taken together, our results provide circumstantial evidence for a cosmologically recent $G$-transition, around 73 million years ago, as a resolution to the Hubble tension. Unlike the proposal of a $G$-transition at the edge of the calibrator step suggested in \cite{Marra:2021fvf} as a resolution to the Hubble tension where the authors assumed that SN standardized peak luminosity $L$ scales in proportion to the Chandrashekhar mass $M_c$, our scenario suggests an \textit{inverse} relationship between $L$ and $M_c$, which is in line with the expectations of~\cite{Wright2018}.

In the future, with a larger calibrator galaxy sample, it might be possible to test more definitively for a transition in the peak SNe magnitudes. Moreover, a careful examination of SNe light curves might show two distinct classes of light curves on either side of this transition which could perhaps help pin down the value of $\Delta G/G$. Another interesting avenue to pursue would be to study the effect of the $G$-transition on TRGB calibrators. TRGB standard luminosities would be expected to transform differently from that of Cepheids under a $G$-transition and thus they would provide a confirmation of the transition distance, and could also pin down the magnitude of the $G$-transition.

Additionally, if SNe~Ia simulations improve to the point where light curves can be reliably predicted, then performing SNe Ia simulations with different values of $G$ could allow for a test of the relationship between $L$ and $M_c$, thus providing an alternative way to test the inverse relationship that we have found here.

We have noted that with a $G$-transition there would also be deviations in the form of the Hubble law for Hubble flow SNe. The modified law at low redshifts (but still above $z_T$) would take the form $d_L(z) = \frac{c z}{H_{0\beta}}(1 + \frac{k}{z} + k)$. We have seen that the $k$ dependent corrections are small for Hubble flow SNe which have a typical redshift $z_\textrm{SN} \sim 0.1$, but we could also in principle test this deviation from the Hubble law with a larger data set of Hubble flow SNe, with a tighter selection criteria on their maximum redshift.

The ``sudden'' $G$-transition that we have proposed in this work is different from the more gradual time variations of $G$ that have been suggested and constrained in the literature (see for e.g.~\cite{Mould:2014iga}) and a number of constraints suggested by these works would not apply. Our hypothesis of a change in $G$ is an idealization, but even if we were to take into account a more gradual change, we do not expect that the variation with time of $G$ is significant at distances larger than 80~Mpc because of the effect on standardization of SNe light curves (see discussion in sec.~\ref{sec:SNeIaG}).

A $G$-transition would not only alter the inferred value of the Hubble constant inferred from the CMB, it could potentially alter the quality of the fits and run into conflicts with observations. In this regard, studies with Planck 2018 CMB data combined with BAO data \cite{Wang:2020bjk, Ballardini:2021evv, Sakr:2021nja} have suggested that a change in the gravitational constant of around 5\% between the present day and in the early universe is allowed at the 2$\sigma$ level. Results from Big Bang Nucleosynthesis (BBN) also allow a 5\% change at 2-$\sigma$ between the BBN and the current era~\cite{Alvey:2019ctk}. Although the cosmological assumptions of these scenarios are different from ours, we find that at face value, the 4\% change in the value of $G$ suggested by our work is not disfavoured by other cosmological data sets.

\section*{Acknowledgements}
We would like to thank Varun Bhalerao for helpful discussions. We also thank the anonymous referee for their insightful comments and criticisms. We acknowledge IUCAA, Pune, India for use of their computational facilities. Ruchika acknowledges financial support from TASP, iniziativa specifica INFN, where a part of the work has been done.

\appendix

\section{Realizing a $G$-transition in a scalar-tensor theory}
\label{sec:scalartensor}
Scalar-tensor theories are one well known mechanism by which the effective gravitational constant becomes dynamical, i.e. it can acquire a time and space variation. The action in a general scalar-tensor theory in the Jordan frame takes the following form~\cite{Faraoni:2004pi},
\begin{equation}
\label{eq:formA}
 S[g_{\mu\nu}, \phi] = \frac{1}{16 \pi} \int d^4x \sqrt{-g}\left [ \phi R - \frac{\omega(\phi)}{\phi} g^{\mu\nu}\nabla_\mu \phi \nabla_\mu \phi - V(\phi) + \mathcal{L}_m (g_{\mu\nu} , \psi_i)\right ] ,
\end{equation}
where $g$ is the usual metric, $R$ is the Ricci scalar constructed form the metric, $\phi$ is the non-minimally coupled scalar field with potential $V(\phi)$, and $\mathcal{L}_m$ is the matter field Lagrangian that depends only upon the metric and the matter fields but is assumed to be independent of $\phi$. The functions $\omega(\phi)$ and $V(\phi)$ can be chosen arbitrarily to give different classes of scalar-tensor theories. The same action as above shows up in different forms in the literature with various redefinitions of the field $\phi$, but for the purposes of our discussion we will assume the form above. Note that Brans-Dicke theory is a special case of this scalar-tensor theory action, with $\omega(\phi) = \omega$, where $\omega$ is reduced to a coupling constant. We will proceed with the more general form that we have written above. The conditions for this general scalar-tensor theory to converge to general relativity are $\omega \rightarrow \infty$ and $\frac{1}{\omega^3} \frac{d\omega}{d\phi} \rightarrow 0$~\cite{Faraoni:2004pi}.

If we compare the scalar-tensor action with the Einstein-Hilbert action of general relativity $ S_{\textrm{EH}} = \frac{1}{16 \pi G_{ \textrm{EH} }} \int d^4x \sqrt{-g} R $, we see that in scalar-tensor theory, we can define an effective gravitational constant $G_{\mathcal{L}}= \frac{1}{\phi(x)}$ which shows up in the Lagrangian as the coefficient of the Ricci scalar in the action given by eq.~\ref{eq:formA}. Here the subscript $\mathcal{L}$ is to remind us that this is an effective gravitational coupling constant that shows up in the Lagrangian. However this coupling is \textit{not} the same as the effective gravitational coupling in the Newtonian limit, which shows up as the coefficient of the inverse square-law force between two test particles.

The effective gravitational coupling between two test masses seen in an experiment will depend on the mass of the scalar degree of freedom $m_\phi$ and the length scale $l$ at which the tests masses are separated. If the effective gravitational force is studied between objects on length scales $l \ll (m_\phi)^{-1}$, then the force will be effectively a standard $1/r^2$ type force with coefficient~\cite{Esposito-Farese:2000pbo} given by,
\begin{equation}
\label{eq:Geffmasslessphi}
G_{\textrm{IS}} = \frac{1}{\phi}\left( \frac{2\omega(\phi) + 4}{2\omega(\phi) +3} \right),
\end{equation}
where the subscript ${\textrm{IS}}$ is to indicate that this is the coefficient of the inverse-square law force in the Newtonian regime. The deviation of $G_{\textrm{IS}}$ away from $G_{\mathcal{L}}$ is due to the extra contribution to the long-range force due to the presence of the nearly massless, long-range scalar.

On the other hand if $l \gg (m_\phi)^{-1}$, then the scalar field is short-ranged and ineffective at these scales. Of course we still get a $1/r^2$ type force but this is purely from the tensor contribution. In this case the effective coefficient of the inverse-square law force is just $G_{\mathcal{L}}$.

Note that in either case of a short-range or long-range scalar force, the field $\phi$ is generically a function of spacetime co-ordinates. \textit{This implies that the effective gravitational coupling can also now acquire a spacetime dependence.}

Various screening mechanisms of scalar-tensor theories have been proposed where the effective mass of the scalar field is environment/density dependent~\cite{Khoury:2003aq,Khoury:2003rn,Hinterbichler:2010es,Brax:2010gi,Babichev:2009ee,VAINSHTEIN1972393}.
We will however proceed with the assumption of a nearly massless scalar field, at least on stellar physics length scales, so that the effective coupling is of the long-range type (with coupling $G_{\textrm{IS}}$).

\vspace{5mm}
Solar-system tests and terrestrial probes strongly constrain such a long-range scalar field. The constraints are  more generally phrased in terms of constraints on the parameters of the so-called parameterized post-Newtonian (PPN) formalism. In any metric theory of gravity, one can parameterize the corrections to the weak-field Newtonian approximation of general relativity as follows,
\begin{eqnarray}
g_{00} &= & - 1 + \frac{2 G_N m}{r c^2 } + 2 \beta^{\textrm{PPN}} \left (\frac{2G_N m}{rc^2}\right)^2, \\
g_{ij} &=& \left( 1 + 2 \gamma^{\textrm{PPN}} \frac{2 G_N m}{r c^2 } \right) \delta_{ij}.
\end{eqnarray}
Here, $\beta^{\textrm{PPN}}$ and $\gamma^{\textrm{PPN}}$ are parameters of the post-Newtonian theory and $G_N$ is the universal gravitational constant with value $G_N= 6.67 \times 10^{-11}$~N-m/kg$^2$ as measured in terrestrial experiments~\cite{10.1093/nsr/nwaa165} which probe the $1/r^2$ law force.

If we have a fundamental (covariant Lagrangian) theory of gravity, we can perturbatively expand the metric in the weak-field limit of a spherically symmetric solution and then map this to the PPN metric. This allows us to identify the parameters of the post-Newtonian theory with the parameters of the Lagrangian.

In standard Einstein gravity, one obtains $\beta^{\textrm{PPN}}=\gamma^{\textrm{PPN}} =1$ and the Lagrangian parameter  $G_{\textrm{EH}}$ appearing in the coefficient of the Ricci scalar is identified with $G_N$.

In scalar-tensor theory, we would identify the present-day local value of the effective gravitational coupling of eq.~\ref{eq:Geffmasslessphi} with the universal gravitational constant, i.e. $G^0_{\textrm{IS}} = G_N$. The parameters $\beta^{\textrm{PPN}}$ and $\gamma^{\textrm{PPN}}$ can be computed in this theory, and in the case where the potential $V = 0$ in eq.~\ref{eq:formA} we obtain the predictions~\cite{Damour:1992we,Damour:1995kt},
\begin{eqnarray}
\gamma^{\textrm{PPN}} &= & \frac{\omega_0 +1 }{\omega_0+2},\\
\beta^{\textrm{PPN}} &=& 1 + \frac{\omega_0^\prime}{(2\omega_0+3)(2\omega_0+4)^2},
\end{eqnarray}
where $\omega_0$ is the present-day value of $\omega(\phi)$ in the solar-system and $\omega^\prime = \frac{d\omega}{d\phi}$.

The parameters $\beta^{\textrm{PPN}}$ and $\gamma^{\textrm{PPN}}$  have been constrained by various solar-system tests, such as the shift of Mercury's perihelion, lunar laser ranging experiments, solar gravitational effects on electromagentic waves etc. (for details, see for example \cite{Uzan:2010pm} and references therein). These tests typically constrain $\beta^{\textrm{PPN}}-1$ and $\gamma^{\textrm{PPN}}-1$ to be less than 1 part in $10^3$ to $10^5$. For the scalar-tensor model this would imply a constraint $\omega_0 > 10^3$ - $10^5$. For such large values of $\omega_0$, we would also find that the present day values of $G_{\textrm{IS}}$ and $G_\mathcal{L}$ and are nearly identical, i.e.  $G^0_{\textrm{IS}}= G_N \simeq G^0_\mathcal{L}$ to the same accuracy as above. However, since solar-system tests only constrain the present day value of $\omega(\phi)$, its value could have been smaller in the past, and thus the difference between $G_{\textrm{IS}}$ and $G_\mathcal{L}$ could be substantial at earlier epochs.

\vspace{5mm}

Let us now turn our attention to the cosmology of scalar-tensor theories. As in Einstein's general relativity, one can also get Friedmann–Robertson–Walker metric (FRW)-like cosmological solutions in a scalar-tensor theory in which the metric takes the usual form,
\begin{equation}
ds^2 = -c^2 dt^2 + a^2(t) \left( dr^2 + r^2 d\Omega^2 \right),
\end{equation}
where we have assumed a spatially flat universe for simplicity. The evolution equations for the scale factor $a(t)$ in the scalar-tensor theory have a different form from the standard Einstein gravity expressions and are given by~\cite{Faraoni:2004pi},
\begin{eqnarray}
\label{eq:FRW1st}
\left ( \frac{\dot{a}}{a} \right )^2 &=& \frac{8 \pi}{3\phi} \rho^{(m)} + \frac{\omega(\phi)}{6} \left(\frac{\dot{\phi}}{\phi} \right)^2 - \frac{\dot{a}}{a} \frac{\dot{\phi}}{\phi} + \frac{V}{6\phi},\\
\label{eq:FRW2st}
\frac{\ddot{a}}{a} &=&  - \frac{4\pi }{3\left ( 1 + \frac{3}{2\omega} \right)} \frac{1}{\phi}
\left [ \left (1 + \frac{3}{\omega} \right ) \rho^{(m)}
 + 3 P^{(m)} \right ] \nonumber
 \\
 && - \frac{\omega}{3} \left(\frac{\dot{\phi}}{\phi} \right)^2 + \frac{\dot{a}}{a} \frac{\dot{\phi}}{\phi} + \frac{1}{2(2 \omega + 3)\phi}\left [ \phi\frac{dV}{d\phi} + \frac{2\omega -3}{6}V  + \frac{d\omega}{d\phi} \dot{\phi}^2 \right ]
\end{eqnarray}

In addition to these equations, we also have the equation of evolution of the scalar field which reduces to,
\begin{equation}
\label{eq:phicosmology}
\ddot{\phi} + \left [ 3 \left( \frac{\dot{a}}{a} \right) + \frac{\dot{\omega}}{2 \omega + 3} \right ] \dot{\phi} = \frac{1}{2\omega +3} \left[ 8 \pi \left(\rho^{(m)} - 3 P^{(m)} \right) - \phi \frac{dV}{d\phi} + 2 V \right].
\end{equation}

In these equations $\rho^{(m)}$ and $P^{(m)}$ are the energy density and pressure of the matter fields (other than that of the gravitational scalar-tensor sector).

This should be compared with the standard evolution of the FRW scale factor in Einstein-Hilbert gravity in the absence of a scalar field,
\begin{eqnarray}
\label{eq:FRW1} \left ( \frac{\dot{a}}{a} \right )^2 &=& \frac{8 \pi G_N}{3} \rho^{(m)},\\
\label{eq:FRW2}  \frac{\ddot{a}}{a} &=& - \frac{4\pi G_N}{3} \left(\rho^{(m)} + 3 P^{(m)} \right ).
\end{eqnarray}
Note that the constant $G_N$ that appears in these equations is the same universal Newtonian gravitational constant that appears as the coefficient of the $1/r^2$ law.

Comparing the FRW equations of scalar-tensor theory (eqs.~\ref{eq:FRW1st} and ~\ref{eq:FRW2st}) with the standard ones of Einstein gravity (eqs.~\ref{eq:FRW1} and ~\ref{eq:FRW2}), we see that there is a limit in which we can obtain the standard FRW evolution equations in a scalar-tensor theory. To obtain this limit we need to take $\phi = \phi_*$ (a constant), $V(\phi_*) \ll \rho^{(m)}, P^{(m)}$, and $\omega \gg 1$. The equations then yield identical expressions for the scale-factor evolution with the replacement of $ G_N \rightarrow\frac{1}{\phi_*}$. Also, as noted earlier, we need the limit of $\omega_0 \gg 1$ in order to be consistent with solar-system tests. In the limit that we are considering, $\omega$, $G_{\textrm{IS}}$, and $G_\mathcal{L}$ are all constant and large, and since $\omega$ is large we have $G_{\textrm{IS}}= G_N \simeq G_\mathcal{L} = \frac{1}{\phi_*}$ in the scalar-tensor theory. Thus, we see that in this limit we do indeed recover both the standard inverse-square law and the standard FRW equations, and the gravitational constant that appears in both places is $\frac{1}{\phi_*} = G_N$. Thus, we see that there is a limit of scalar-tensor theory where we identically recover all the standard predictions of general relativity, both for cosmology and in the Newtonian regime.

How do we ensure that the limit that we considered above is self-consistent? One needs to check that we can find an $\omega$ and $V$ such that $\phi = \phi_*$ is also a solution of the $\phi$ evolution equation (eq.~\ref{eq:phicosmology}). It is easy to see that in the limit of large $\omega$ that this is indeed a valid solution to this equation.

\vspace{5mm}
Now, we would like to consider a different solution to the $\phi$ equation of motion in order to account for a $G$-transition in the scalar-tensor model. We will assume that the scalar functions $\omega (\phi)$ and $V(\phi)$ can be chosen so as to yield a solution to the $\phi$ evolution equation (eq.~\ref{eq:phicosmology}) which takes the form,
\begin{equation}
\label{eq:phievolution}
\phi(x,t) = \begin{cases}
 \phi_{\alpha} & \text{ for } t \geq t_T, \\
 \phi_{\beta} & \text{ for } t < t_T,
 \end{cases}
\end{equation}
where $t_T$ is a transition time, and $\phi_\alpha$, $\phi_\beta$ are constants. Thus, for very early times (smaller than the transition time $t_T$) the FRW equations of the scalar-tensor theory look identical to that of the standard cosmological evolution with effective gravitational constant in the FRW equations $G_\textrm{eff} =\frac{1}{\phi_\beta}$, and similarly for very late times the equations once again look similar to that of the standard cosmology, but this time with coefficient $G_\textrm{eff} =\frac{1}{\phi_\alpha}$.

This hypothesis effectively arranges for a cosmological $G$-transition. Such a discrete change in $\phi$ would show up not just in the FRW equations, but also as a change in the coefficient $G_\textrm{IS}$ of the inverse-square law at early-times and at late times. However, it is only the late time value of $\phi$ that is related to the Newtonian gravitational constant which is measured in the present-day solar-system and thus $G_N =\frac{1}{\phi_\alpha}$ in this scenario.

We have assumed for simplicity in eq.~\ref{eq:phievolution} that the transition in the value of $\phi$ and hence the gravitational coupling is instantaneous, but one could also assume a finite width for the transition at the cost of introducing some additional parameters into the hypothesis.

The key model building challenge to ensure the self-consistency of this solution is to ensure that the potential $V$ and the function $\omega$ are such that they lead to the dynamics of $\phi$ that yield such a transition, while simultaneously ensuring that the model is consistent with solar-system tests (which can be ensured for $\omega_0 >> 1$). In addition the dynamics of $\phi$ must be such that it does not significantly alter the standard cosmology -- other than through the $G$-transition -- in order to avoid conflict with existing cosmological observations. This means that the scale-factor evolution must be similar to that of the standard FRW evolution and the scalar potential should contribute negligibly to the energy-density, so we need to assume that $V(\phi_A), V(\phi_B) \ll \rho^{(m)}, P^{(m)}$. We also assume that the inhomogeneities in $\phi$ can be neglected, so that they do not affect the evolution of the other density perturbations.

The difficulties with arranging for a $G$-transition that satisfy these criteria are apparent if for example one attempts to consider the simple guess that $\omega \gg 1$ at all times. This would make the dynamics of $\phi$ nearly insensitive to the form of the potential in eq.~\ref{eq:phicosmology} and would yield a solution where $\phi$ is a constant and thus yield no $G$-transition. To find a self-consistent solution with a $G$-transition then, one might need to consider the possibility that $\omega \sim 1$ at early times, or perhaps one could consider a non-minimal scalar sector that could lead to the appropriate dynamics.

Demonstrating the existence of such a self-consistent scenario for a $G$-transition and constructing a toy model for it are beyond the scope of the present work. We will assume that these challenges can be overcome and we leave the details of such model building (assuming that it can be done) to future work.

\section{Effect of a $G$-transition on the value of the Hubble constant inferred from the CMB}
\label{sec:appendixB}
We first discuss the effect of a $G$-transition on the sound horizon $r_s(z_*)$ and then we discuss the effect on the inferred value of the Hubble constant from the CMB.
\subsection{Effect of a $G$-transition on the sound horizon $r_s(z_*)$}
\label{sec:appendixB1}
The physical sound horizon $r_s(z_*)$ is given by,
\begin{equation}
\label{eq:rszstar}
r_s(z_*) = \frac{1}{1+z_*} \int_{z_*}^\infty \frac{dz}{H(z)} c_s(z),
\end{equation}
where $H(z)$ is the Hubble rate and $c_s(z)$ is the sound speed of the baryon-photon fluid. In the pre-recombination era the Hubble rate is given in terms of the present day matter density $\rho^{(0)}_m$ and radiation density $\rho^{(0)}_\textrm{rad}$ as,
\begin{eqnarray}
H^2(z)&\simeq& \frac{8\pi G}{3} \left[ \rho^{(0)}_m (1+z)^3 + \rho^{(0)}_\textrm{rad}(1+z)^4 \right], \\
&=& H^2_{100}  \omega_m (1+z)^3 + \frac{8\pi G}{3} \rho^{(0)}_\textrm{rad} (1+z)^4,
\end{eqnarray}
where in the second line we have rewritten the matter density in terms of $\omega_m$, and we have defined the constant $H_{100} = 100$~\Hunit. $\rho^{(0)}_\textrm{rad}$ is the total present day radiation density which is known from the observed CMB black-body temperature $T = 2.725$~K and the number of neutrino species (which we assume is 3), and its value is $\rho^{(0)}_\textrm{rad} = 8.09 \times 10^{-34}$~g/cc.

We can write the sound speed $c_s(z) = \sqrt{\frac{1}{3(1+R)}}$, where $R= \frac{3}{4}\frac{ \rho_b}{\rho_\gamma}$ in terms of the independent parameters as,
\begin{equation}
c^2_s(z) = \frac{c^2}{3} \frac{1}{\left(1 + \frac{3}{4} \frac{\omega_b H_{100}^2}{\frac{8\pi G}{3}\rho^{(0)}_\gamma}\frac{1}{(1+z)} \right)},
\end{equation}
where $\omega_b$ is the Hubble-scaled baryon density fraction and $\rho^{(0)}_\gamma = 4.813\times10^{-34}$~g/cc is the present day energy density of photons.
We can substitute the expressions for $c_s(z)$ and $H(z)$ into eq.~\ref{eq:rszstar} and numerically integrate to obtain,
\begin{equation}
r_s(z_*) = \frac{143.6~\textrm{Mpc}}{1+z_*},
\end{equation}
where have taken $z_* = 1089, \omega_m = 0.144$, and  $\omega_b =0.0224$ based on the Planck cosmological fits~\cite{p2020}. The value of 143.6~\textrm{Mpc} represents the size of the comoving sound horizon.

Next, we would like to see the effect of a change in $G$ on $r_s(z_*)$. First, with a little work, we can easily write down the first-order change in $H(z)$ and $c_s(z)$ by differentiating their respective expressions with respect to $G$ to obtain,
\begin{eqnarray}
\frac{\Delta H(z)}{H(z)} &=& \frac{\Delta G}{G} \left[\frac{1}{2} \frac{\frac{8\pi G}{3} \rho^{(0)}_\textrm{rad} (1+z)^4}{ H^2_{100}  \omega_m (1+z)^3 + \frac{8\pi G}{3} \rho^{(0)}_\textrm{rad} (1+z)^4} \right ],  \\
\frac{\Delta c_s(z)}{c_s(z)} &=& \frac{\Delta G}{G} \left[\frac{3}{8} \frac{\omega_b H_{100}^2}{\frac{8\pi G}{3}\rho^{(0)}_\gamma}\frac{1}{(1+z)} \frac{c_s^2(z)}{c^2/3} \right ].
\end{eqnarray}
Now the change in $r_s(z_*)$ is given by,
\begin{equation}
\label{eq:deltarszstar}
\Delta r_s(z_*) = \frac{1}{1+z_*} \int_{z_*}^\infty \frac{dz}{H(z)} c_s(z) \left[ -\frac{\Delta H(z)}{H(z)} + \frac{\Delta c_s(z)}{c_s(z)}  \right ].
\end{equation}

Pulling out a factor of $\Delta G/G$ allows us to evaluate this integral numerically to obtain,
\begin{equation}
\label{eq:deltarszstarnum}
\Delta r_s(z_*) = -\frac{23.1}{1+z_*} \left ( \frac{\Delta G}{G} \right )~\textrm{Mpc}.
\end{equation}

Thus, the fractional change in the sound horizon under a change in $G$ is given by,
\begin{equation}
\label{eq:deltarszstarnum}
\frac{\Delta r_s(z_*)}{ r_s(z_*)} = -0.16 \left( \frac{\Delta G}{G} \right ).
\end{equation}

\subsection{Effect of a $G$-transition on the inferred value of the Hubble constant}
\label{sec:appendixB2}
In sec.~\ref{sec:CMBinference}, we have argued that in a $G$-transition cosmology, the angular diameter distance is modified to,
\begin{equation}
\label{eq:dAzstar3}
d_A(z_*) \simeq \frac{1}{H_{100}}\frac{1}{1+z_*}  \bigints_{0}^{z_*} \frac{1}{\left[ \omega_m (1+z)^3+   \left(\frac{H^2_{0\beta}}{H^2_{100}}- \omega_m \right) \right]^{1/2}} dz,
\end{equation}
where the difference from the standard cosmology is only the replacement of $H_0$ by $H_{0\beta}$. This equation provides a relationship between $d_A(z_*)$ and $H_{0\beta}$ for a given value of $\omega_m$.

The Hubble constant can be calculated by inverting this relationship and setting $d_A(z_*) = r_s(z_*)/\theta_*$, where $\theta_*$ is the observed angular size of the first peak in the CMB.
Now assuming that the fit to $\omega_m$ is left unchanged, we see that the effect of a $G$-transition is that it will lead to a change in $r_s(z_*)$ which will correspondingly lead to a change in the value of $H_{0\beta}$, changing its value away from the value of $H_0$ as inferred from the CMB assuming the standard cosmology. We can evaluate the change in $H_{0}$ (or $H_{0\beta}$) by differentiating eq.~\ref{eq:dAzstar3} to obtain the first order change,
\begin{equation}
\frac{\Delta H_{0}}{H_{0}} = \frac{1}{\frac{d \textrm{Log} \, d_A(z_*)}{d \textrm{Log}\, H_{0\beta}}}  \frac{\Delta r_s(z_*)}{r_s(z_*)}.
\end{equation}
The derivative $\frac{d \textrm{Log} \, d_A(z_*)}{d \textrm{Log}\, H_{0\beta}}$ can be evaluated numerically and gives a value of -0.19, when setting $H_{0\beta}/H_{100} \simeq H_{0}/H_{100} = 0.67$ and $\omega_m = 0.144$.
Thus we find,
\begin{equation}
\frac{\Delta H_{0}}{H_{0}} \simeq -\frac{1}{0.19}  \frac{\Delta r_s(z_*)}{r_s(z_*)} \simeq +0.83 \frac{\Delta G}{G},
\end{equation}
where $H_{0\beta} = H_{0} + \Delta H_{0}$.

\bibliographystyle{JHEP.bst}
\bibliography{references}

\end{document}